\documentclass[12pt]{iopart}
\usepackage{iopams}
\usepackage{graphicx}
\usepackage{bm}
\usepackage{dcolumn}
\usepackage{color}
\usepackage[latin9]{inputenc}
\usepackage{url}
\usepackage{subfigure}
\usepackage{float}
\usepackage{longtable}
\usepackage{epstopdf}

\sloppypar

\newcommand{\raiseentry}[1]{\smash{\raise 0.7 em \hbox{#1}}}

\newenvironment{equationarray*}
{\arraycolsep 0.14 em
\begin{eqnarray*}}
{\end{eqnarray*}}

\begin{document}

\title[Gravitational Wave Signals Contaminated by Transient Detector Noise Glitches]{Parameter Estimation and Model Selection of Gravitational Wave Signals Contaminated by Transient Detector Noise Glitches}

\author{Jade Powell$^1$}

\address{$^1$ OzGrav, Swinburne University of Technology, Hawthorn, VIC 3122, Australia }

\date{\today}


\begin{abstract}
The number of astrophysical sources detected by Advanced LIGO and Virgo is expected to increase as the detectors approach their design sensitivity. Gravitational wave detectors are also sensitive to transient noise sources created by the environment and the detector, known as `glitches'. As the rate of astrophysical sources increases, the probability that a signal will occur at the same time as a glitch also increases. This has occurred previously in the gravitational wave binary neutron star detection GW170817. In the case of GW170817, the glitch in the Livingston detector was easy to identify, and much shorter than the total duration of the signal, making it possible for the glitch to be removed. In this paper, we examine the effect of glitches on the measurement of signal parameters and Bayes factors used for model selection for much more difficult cases, where it may not be possible to determine that the glitch is present or to remove it. We include binary black holes similar to current detections, sine Gaussian bursts, and core-collapse supernovae. We find that the worst effects occur when the glitch is coincident with the signal maximum, and the signal to noise ratio (SNR) of the glitch is larger than the signal SNR. We have shown that for accurate parameter estimation of future gravitational wave signals it will be essential to develop further methods to either remove or reduce the effect of the glitches.
\end{abstract}

\section{Introduction}
\label{sec:intro}

The Advanced LIGO (aLIGO) \cite{aLIGO} and Virgo (AdVirgo) \cite{AdVirgo} gravitational wave detectors have made the first direct detections of gravitational waves from binary neutron stars (BNS) and binary black holes (BBH) \cite{PhysRevLett.116.061102, PhysRevLett.116.241103, PhysRevLett.118.221101, 2017arXiv171105578T, PhysRevLett.119.141101, PhysRevLett.119.161101}. When the advanced detectors begin their third observing run the number of detections of binary sources is expected to increase \cite{2016ApJ...833L...1A}. The detectors may also detect gravitational waves from other transient sources, such as, core-collapse supernovae (CCSNe) \cite{2016arXiv160501785A}, gravitational wave orphan memory \cite{2017arXiv170201759M}, cosmic strings \cite{2005PhRvD..71f3510D}, or a totally unknown astrophysical source. 

Estimating parameters of gravitational wave detections is essential for understanding the physics of the source. For binary sources, the parameters include the mass, spin, sky position, distance and inclination. Estimating the sky position and distance allows electromagnetic follow up of the gravitational wave signal. This resulted in the first confident joint electromagnetic and gravitational wave detection from GW170817 and GRB170817A \cite{PhysRevLett.119.161101, 2041-8205-848-2-L12}. Measuring the mass and spin of a population of binary sources can help us understand binary formation \cite{0264-9381-27-11-114007, 2017Natur.548..426F, 2017MNRAS.471.2801S, 2015ApJ...810...58S, 0264-9381-34-3-03LT01, 2017PhRvD..95l4046G, 2018arXiv180102699T}. For a detection of a CCSN signal, parameter estimation and model selection could tell us about the rotation of the proto-neutron star \cite{2015MNRAS.450..414F, 2015ApJ...811...86Y, 2014PhRvD..90d4001A}, the equation of state \cite{2017PhRvD..95f3019R}, turbulent features such as convection and the standing accretion shock instability (SASI) \cite{2017arXiv170801920T}, and the explosion mechanism of the source \cite{2016PhRvD..94l3012P, 2017PhRvD..96l3013P}.  

Detecting gravitational wave sources is difficult due to the non-stationary and non-Gaussian nature of the detector noise. Improvements to the detector noise have been made in previous studies by improving measurements of the power spectral density (PSD) \cite{2015PhRvD..92f4011E, 2015PhRvD..91h4034L}. The data contains short duration transient noise artefacts, called \textit{glitches}, that can reduce the duty cycle of the instruments and limit the sensitivity of gravitational wave searches. Previous studies have implemented methods to reject glitches by analysing data coherently and incoherently and comparing the results \cite{2015CQGra..32m5012C}.

During the first aLIGO Observing Run (O1), $10^{6}$ glitches above a signal to noise ratio (SNR) 6 were observed in 51.5 days of data \cite{2018CQGra..35f5010A}. The high rate of glitches means that there is a high probability that future gravitational wave detections may occur at the same time as a glitch. This has already occurred during the detection of the BNS signal GW170817 \cite{PhysRevLett.119.161101}. To prevent errors in the analysis of the signal, two different methods were used to remove the glitch from the data. The first method, known as \textit{gating}, applied an inverse Tukey window to zero out the data around the time of the glitch \cite{2016CQGra..33u5004U}. The second method, which was applied before parameter estimation of the signal, involved subtracting the waveform of the glitch that was reconstructed with sine Gaussian wavelets \cite{2015CQGra..32m5012C}. However, these methods have currently not been tested on a wider range of signals and glitches. 

Subtracting the glitch from the GW170817 detection was possible because it could be determined easily in this case that both a signal and glitch were present. This is partly due to the fact that binary signals have a well understood characteristic chirp shape and the signal had a loud SNR. For other types of short duration gravitational wave signals, where the shape is less understood or completely unknown, it may be more difficult to determine if part of the reconstructed gravitational waveform is due to a glitch. The aLIGO detectors contain thousands of instrument and environmental monitors that produce data called auxiliary channels, which are not sensitive to gravitational waves and can be used to \textit{veto} glitches in the detectors. The glitch had a very large amplitude and was visible in auxiliary channels of aLIGO data that are not sensitive to gravitational wave detections. The auxiliary channels allowed us to determine that the GW170817 glitch was produced by a saturation in the digital-to-analog converter of the feedback signal controlling the position of the test masses \cite{PhysRevLett.119.161101}. However, certain glitch types occur only in data that is sensitive to gravitational waves, which would make it much more difficult to determine if the glitch is part of the signal or a noise artefact. In the case of GW170817, the glitch was much shorter in duration than the length of the signal, therefore, gating could be applied without completely removing the signal from one detector. In future detections, if a glitch contaminates a shorter duration signal, such as a high mass BBH or a short duration burst, it may not be possible to gate the glitch without removing the entire gravitational wave signal.

In this study, we aim to determine the effects of different types of aLIGO glitches on the estimated parameters of multiple types of gravitational wave signals that are too short in duration for gating to be applied. We select glitches from the first aLIGO Observing Run (O1), which were still present during the second Observing Run (O2), and have a high probability of still occurring in future observing runs. We use short duration BBH signals, as they are the most common source for ground based detectors, and sine Gaussian signals to examine the effects on short duration burst sources. The signal morphology of a gravitational wave burst is not understood well enough to produce a matched filter template bank due to computational expense or unknown astrophysics. Therefore, it is not possible to have a signal model that will be an exact match for most potential future burst gravitational wave detections, and typically a sine Gaussian signal model is used \cite{essick:15, 2016PhRvD..93d2004K}. As we know that for any burst gravitational wave signal there will be a mis-match between the signal and model, we use a third signal type to examine the effects of glitches when there is a mis-match between the signal and model. For this third type of signal we use supernova signals analysed with a sine Gaussian signal model. 

In Section \ref{sec:paramest}, we describe the parameter estimation and model selection tools used in this study. In Section \ref{sec:analysis}, we outline the analysis and provide a description of the glitches and gravitational wave signals. The results are described in detail in Section \ref{sec:results_bbh} for BBH signals, Section \ref{sec:results_sg} for sine Gaussian burst signals, and Section \ref{sec:results_sn} for supernova signals with a mis-match between the signal and model template. A discussion is given in Section \ref{sec:discussion}.  

\section{Parameter Estimation}
\label{sec:paramest}

Bayesian data analysis tools are used for parameter estimation and model selection of gravitational wave signals. In this study, we use the parameter estimation software library \texttt{LALInference} \cite{2015PhRvD..91d2003V}.  

For a given model $M$, data $d$, and set of parameters $\theta$, Bayes theorem is given by
\begin{equation} 
p(\theta |d,M) = \frac{p(\theta|M) p(d|\theta,M)}{ p(d|M) }\,,
\end{equation}
where $p(\theta|M)$ is the prior that represents what is known about the parameters before any analysis of the data, and $p(d|\theta,M)$ is the likelihood. An odds ratio can be calculated to distinguish between two different models as
\begin{equation}
O_{i,j} = \frac{p(M_{i})}{p(M_{j})} \frac{p(d|M_{i})}{p(d|M_{j})} = \frac{p(M_{i})}{p(M_{j})} B_{ij}\,,
\end{equation}
where $M_{i}$ and $M_{j}$ are the two competing models. If each model has the same prior then the odds ratio is equivalent to the Bayes factor given as
\begin{equation}
B_{ij} = \frac{p(d|M_{i})}{p(d|M_{j})},
\end{equation}
where $p(d|M_{i})$ is the evidence for model $M_{i}$, and $p(d|M_{j})$ is the evidence for model $M_{j}$. In our case, model $M_{i}$ is the data contains noise and a signal, and model $M_{j}$ is that the data contains only noise. The evidence is given by the likelihood multiplied by the prior integrated over all parameter values
\begin{equation}
p(d|M) = \int_{\theta} p(\theta|M) p(d|\theta, M) d\theta.
\end{equation}
Gravitational wave signals typically have a high number of parameters that make the integral computationally challenging. In this study, the evidence integral is solved using nested sampling \cite{skilling:04}. The posterior probability density functions for each parameter are found by marginalizing over all but one or two of the parameters.
The likelihood is given by 
\begin{equation}
\mathcal{L} = \exp \sum_{i} \left[ \frac{-2|\tilde{h}_{i}-\tilde{d}_{i}|^2}{tS_{n}(f_{i})} -\frac{1}{2} \log(\pi t S_{n}(f_{i})/2) \right]
\end{equation}
where $S_{n}$ is the noise power spectral density (PSD), $t$ is the duration of data analysed, $\tilde{h}$ is the Fourier transform of the gravitational wave signal and $\tilde{d}$ is the discrete Fourier transform of $d$. For multiple gravitational wave detectors the likelihood becomes
\begin{equation}
\mathcal{L}_{H,L,V} = \prod_{i\epsilon(H,L,V)} \mathcal{L}_{i} 
\end{equation}

The signal model used for parameter estimation and model selection depends on the type of gravitational wave signal that is being analyzed. For BBH signals, we use \texttt{IMRPhenomPv2} waveforms, a standard phenomenological precessing waveform family \cite{2016PhRvD..93d4006H, 2016PhRvD..93d4007K, 2014PhRvL.113o1101H}. Running \texttt{LALInference} with an \texttt{IMRPhenomPv2} signal model produces a signal vs. noise Bayes factor and posterior distributions on several signal parameters. The parameters typically include the chirp mass given by,
\begin{equation}
\mathcal{M} = \frac{ (m_{1}m_{2})^{\frac{3}{5}} }{ (m_{1} + m_{2})^{\frac{1}{5}} }, 
\end{equation}
where $m_{1}$ and $m_{2}$ are the component masses, as well as the mass ratio, the spin parameters, the distance, the inclination and the sky position of the source. 
 
The Bayes factors are important as they can be used as a search statistic \cite{2017arXiv171200688S}. The mass is important for population studies \cite{2015ApJ...810...58S, 2018arXiv180102699T}. The distance can aid in finding counterparts to the source \cite{2016ApJ...829L..15S}, it can be used to measure the Hubble constant \cite{2017Natur.551...85A}, and it can help to determine if the properties of populations of sources vary with distance.

For the analysis of burst signals with \texttt{LALInference} a sine Gaussian signal model is used, as the exact waveform of gravitational wave burst signals is unknown \cite{essick:15, lynch:15}. The sine Gaussian signal model is defined as,
\begin{equation}
	h_{\times}(t)= h_{0} \sin[2\pi f_{0} (t-t_{0})]e^{-(t-t_{0})^{2}/2\tau^{2}}\,,
	\label{eqn:SGeqn1}
\end{equation}
\begin{equation}
	h_{+}(t)= h_{0} \cos[2\pi f_{0} (t-t_{0})]e^{-(t-t_{0})^{2}/2\tau^{2}}\,,
	\label{eqn:SGeqn2}
\end{equation}
where $\tau$ is the signal duration given by $\tau=Q/\sqrt{2}\pi f_0$, $f_0$ is the central frequency, $Q$ is the quality factor, $t_{0}$ is the GPS time at the centre of the sine Gaussian, and $h_{0}=\hbox{hrss}/\sqrt{\tau}$, where $\hbox{hrss}$ is the root sum squared amplitude of the transient. The burst version of \texttt{LALInference} can produce posterior distributions for $Q$, $f_0$, hrss, and the sky position of the source, as well as signal vs. noise Bayes factors. 
In burst searches, the Bayes factors are used as a search statistic when combined with a trigger generator \cite{lynch:15}. As there is currently no complete set of astrophysical signal models for the majority of burst signals, it is currently not possible to directly measure parameters such as the mass of the source. However, for a burst source such as a supernovae, measurements of the frequency, duration, and hrss will allow us to make comparisons with values predicted in simulation studies \cite{2017arXiv170801920T}, which may allow us to learn about the astrophysics of the source.  

\begin{figure}[!t]
\centering
\includegraphics[width=0.35\textwidth,height=4.5cm]{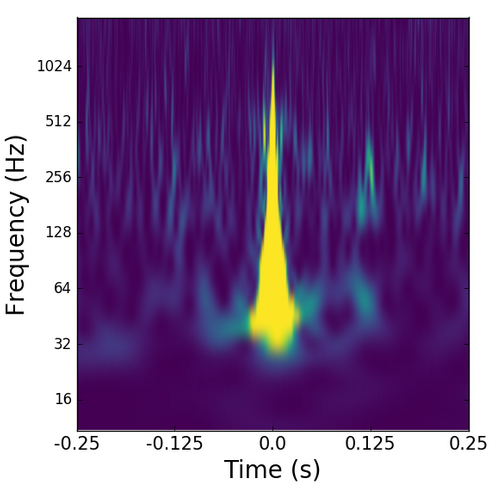}
\includegraphics[width=0.31\textwidth,height=4.5cm]{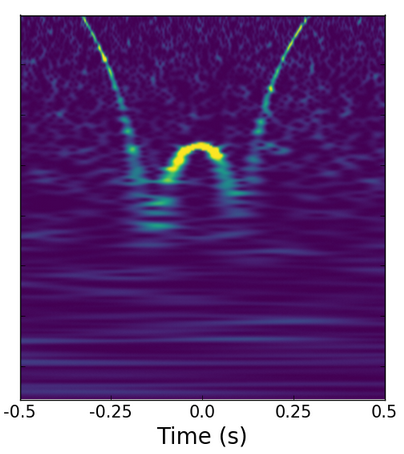}
\includegraphics[width=0.31\textwidth,height=4.5cm]{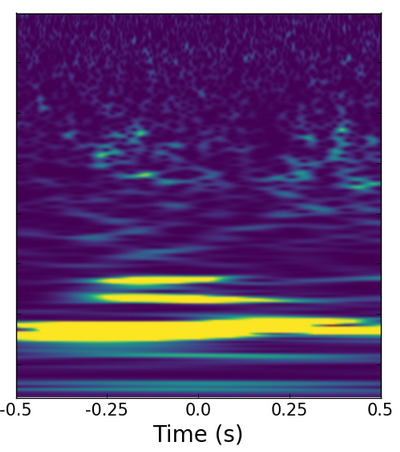}
\caption{A spectrogram of each type of of glitch used in this study. Lighter colours show higher signal energy. (Left) A glitch known as a \textit{blip} characterised by it's tear drop shape in a spectrogram. (Middle) A glitch known as a \textit{whistle}, as this is what they sound like if you listen to the detector noise. (Right) A \textit{scattered light} glitch created by laser light scattering in the gravitational wave detector. Images taken from Gravity Spy \cite{gravityspy}. }
\label{fig:glitches}
\end{figure}

\section{Analysis}
\label{sec:analysis}

To determine how well we can measure source parameters when glitches are present, we use data taken during O1. Simulated gravitational wave signals are added to both the Livingston (L1) detector and the Hanford (H1) detector O1 data at increasing time offsets from three different types of glitches. We use glitches that occur in the L1 detector only at the same time as good quality data in the Hanford H1 detector and data from both detectors is analysed. All of the glitches are identified using glitch classification techniques \cite{gravityspy, 2017CQGra..34c4002P, 2015CQGra..32u5012P}. We inject the signals directly on top of the glitches, 0.1\,s away from the glitches, and 0.2\,s away from the glitches. We do not include the AdVirgo detector, as we use glitches which occurred when the AdVirgo detector was offline for upgrades.


\subsection{Glitches}
\label{sec:glitches}

In this study, we select three of the most common glitches that occur in the aLIGO detectors. The three different glitch types are known as blips, whistles and scattered light. A spectrogram of each glitch type is shown in Figure \ref{fig:glitches}. We select 50 glitches of each type from the L1 aLIGO detector that occur at the same time as good quality data in the aLIGO H1 detector. All the glitches used in this study have an SNR that is between 10 and 20. These values are used as they allow the glitches to be large enough that they can be detected, but not so large that they will be instantly `vetoed' as not being real gravitational wave signals. 

Glitches created from light scattering in the detector are often long duration and low frequency. Scattered light glitches occur when a small fraction of the laser beam light is scattered by imperfections in the mirrors of the detectors and then recombines with the main beam. The 50 scattered light glitches considered in this study have frequencies less than 40\,Hz and average duration's of $\sim 2\,s$. 

The blip glitch is the most common type of glitch that occurs in both of the aLIGO detectors. Therefore, it is possible that this type of glitch will overlap with an astrophysical gravitational wave signal in the future. Their peak frequency is at a few hundred Hz, which is the most sensitive frequency range of the advanced gravitational wave detectors. This results in blip glitches limiting the background sensitivity of searches for astrophysical transient sources. They appear as a short duration $(\sim 0.1\,ms)$ spike in the gravitational wave time series and can occasionally look chirpy in a spectrogram. The source of blip glitches is currently unknown, and is an active area of investigation \cite{nuttall:15}. The blip glitches are not present in any of the auxiliary channels, making it very difficult to gain clues about their origin. 

The whistle glitches, sometimes called radio frequency beat notes, have a much longer duration and a higher frequency than the blip glitches. They often appear to have a `v' or `w' shape in a spectrogram. Some explanation of the origin of the whistle glitches is given in Ref. \cite{nuttall:15}, but they have not yet been eliminated from the detector data. The 50 whistles considered in this study have an average duration of $\sim 0.7$\,s, and an average peak frequency of $\sim 800$\,Hz.


\subsection{Gravitational Wave Signals}
\label{sec:signals}

\begin{figure}[!t]
\centering
\includegraphics[width=0.32\textwidth]{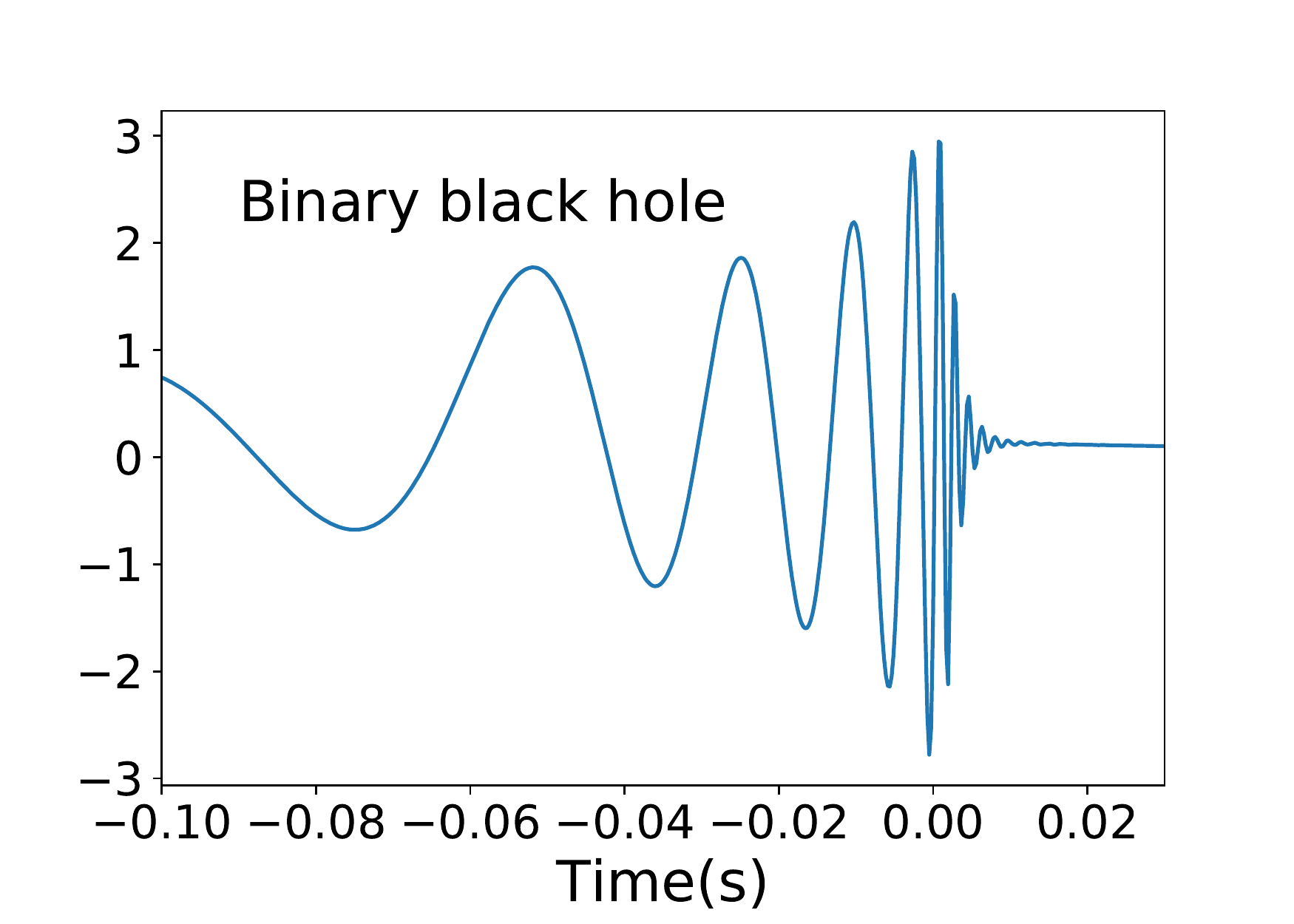}
\includegraphics[width=0.32\textwidth]{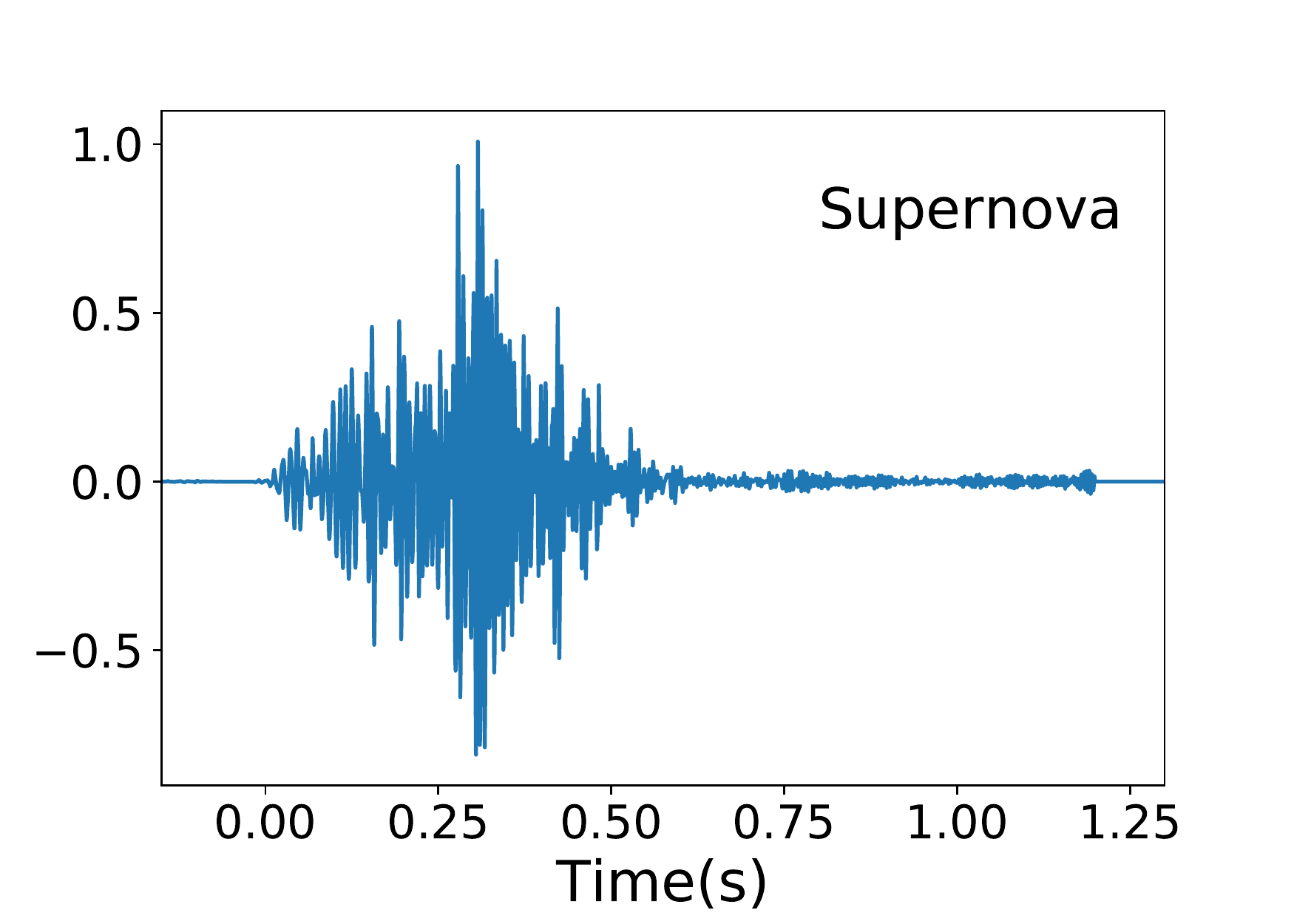}
\includegraphics[width=0.32\textwidth]{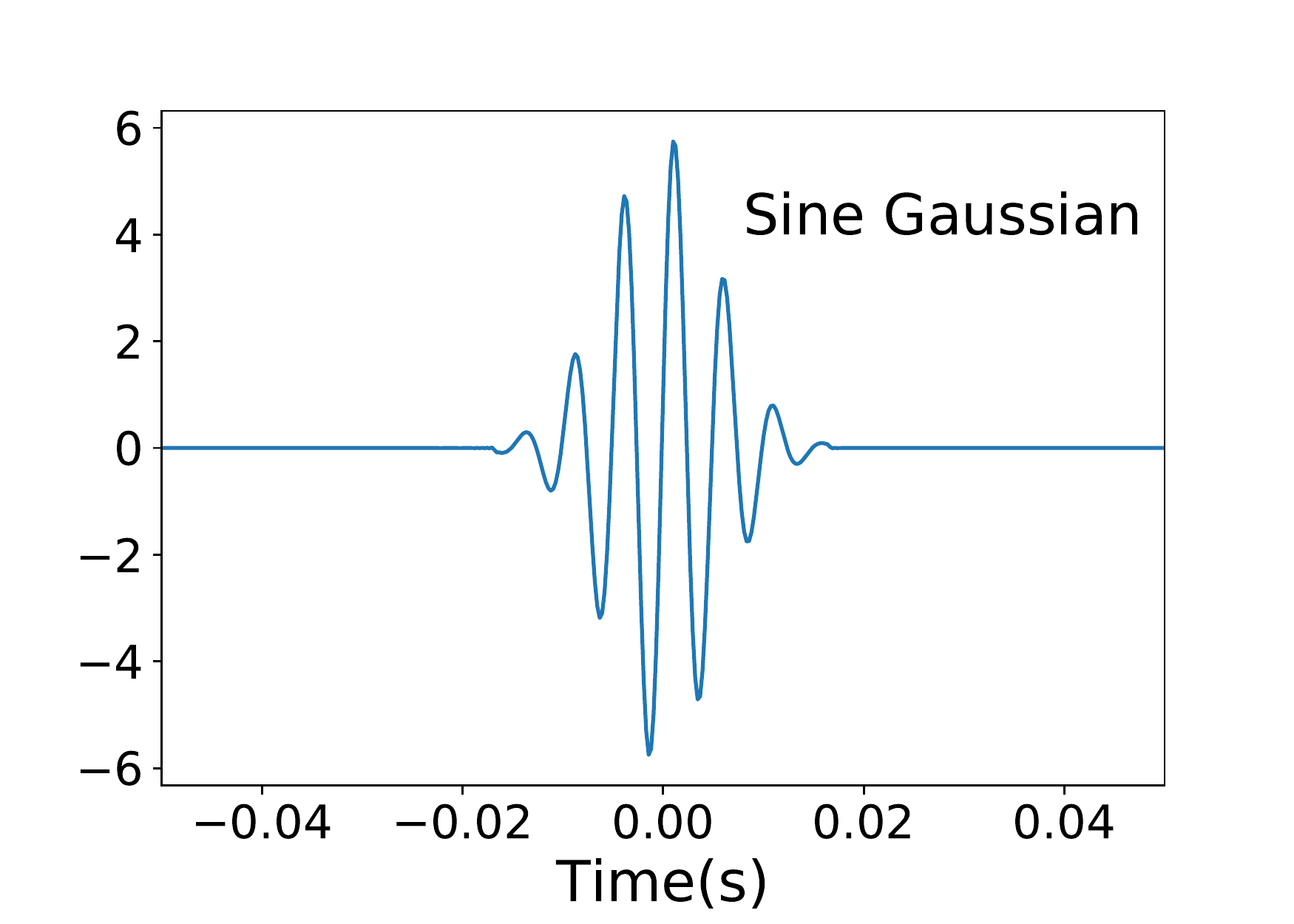}
\caption{The waveforms of the three short duration gravitational wave signals considered in this study. From left to right: a binary black hole, a supernova signal simulated from a $15M_{\odot}$ progenitor star \cite{mueller:e12}, and a sine Gaussian with a frequency of 200\,Hz.}
\label{fig:signals}
\end{figure}

An example of the three different short duration gravitational wave signals used in this study are shown in Figure \ref{fig:signals}. We inject all of the signals exactly on top of glitches, 0.1\,s away from the glitches, and 0.2\,s away from the glitches. All of the signals in this paper have an SNR less than 35. 

The first type of signal is BBH systems with chirp masses in the range $24-31\,M_{\odot}$, and distances in the range 139-700\,Mpc. The signals are distributed uniformly on the sky. All of the spins of the signals are set to zero as we do not consider spin parameters in this study. We calculate signal vs. noise log Bayes factors, and we examine the effect the glitches have on the chirp mass and distance posteriors. 

To represent an unknown burst signal, we use sine Gaussian signals. The signals have a frequency of 200\,Hz, a Q of 5, and a hrss of $8.8\times 10^{-23}$, and are distributed uniformly on the sky. We calculate the signal vs. noise log Bayes factors, and we examine the effect the glitches have on the frequency and hrss posterior distributions. 

To examine if the effect of glitches is worse when there is a mis-match between the signal and the model, we use one supernova signal simulated with a $15\,M_{\odot}$ progenitor star, referred to as model L15 in \cite{mueller:e12}. The signals have a duration of 1.2\,s with all of the detectable parts of the signal within 0.6\,s. The signals have a broad frequency range with a peak at $\sim 100$\,Hz. The signals are injected at distances of 1\,kpc and 0.5\,kpc at the position of the center of the Galaxy. As for the sine Gaussian signals, we calculate the signal vs. noise log Bayes factors with the sine Gaussian signal model, and we examine the effect the glitches have on the duration and hrss posterior distributions. We vary the difference in SNR between all of the different signals and glitches.
 
All of the signals used in this study have a large enough SNR that we expect they would be detected by the searches. The blip glitches limit the sensitivity of gravitational wave searches as they produce large tails in the estimation of the gravitational wave background. However, an individual blip glitch overlapping with a signal should not prevent the signal from being detected by the search, as it is currently not possible to produce a veto for data containing blip glitches. Data containing whistle glitches and scattered light can be vetoed as these glitch types appear in instrument and environment monitors surrounding the detector. However, the loudest background events in searches are checked carefully by the LIGO team to ensure a real signal is not rejected. So it is expect that the signals used in this study that overlap with whistles and scattered light may have some delay in their detection, as was the case for GW170817. However, in the case of GW170817, even with the L1 data initially being vetoed as bad quality, the signal was still detected quickly enough for a detection of the signals electromagnetic counterpart. 



\section{Binary black hole results}
\label{sec:results_bbh}

\begin{figure}[!t]
\centering
\includegraphics[width=0.49\textwidth,height=4.5cm]{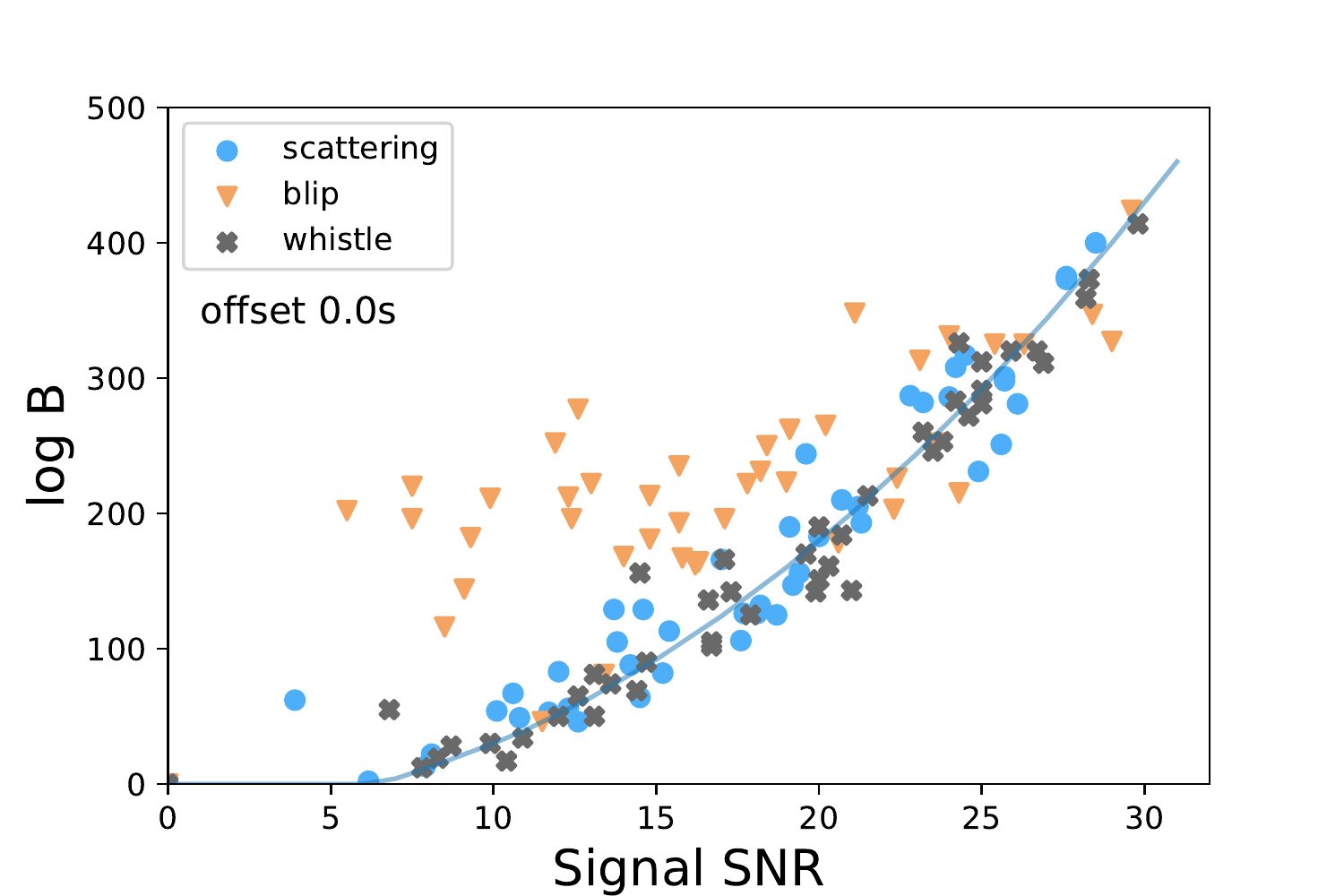}
\includegraphics[width=0.49\textwidth,height=4.5cm]{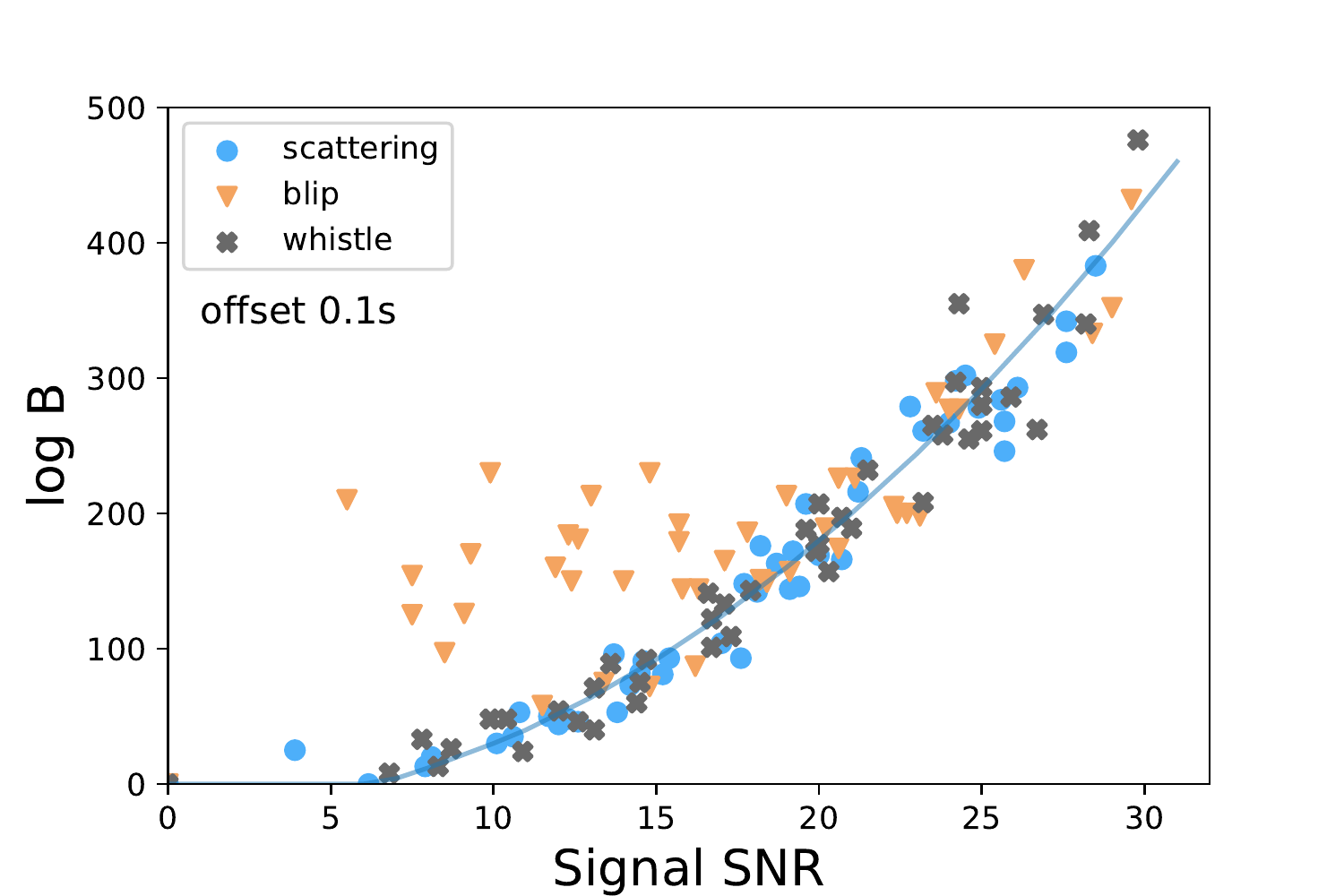}
\includegraphics[width=0.49\textwidth,height=4.5cm]{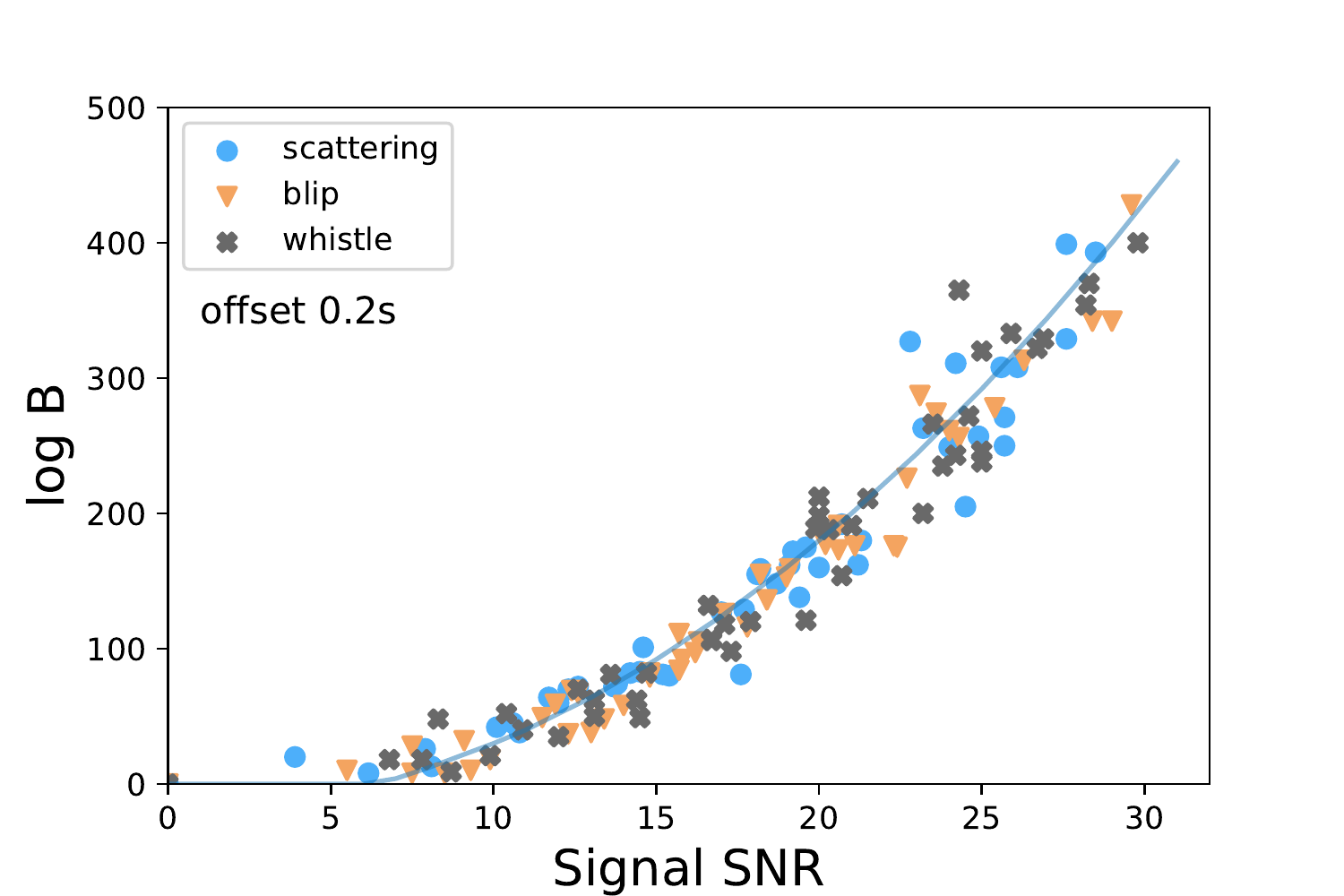}
\includegraphics[width=0.49\textwidth,height=4.5cm]{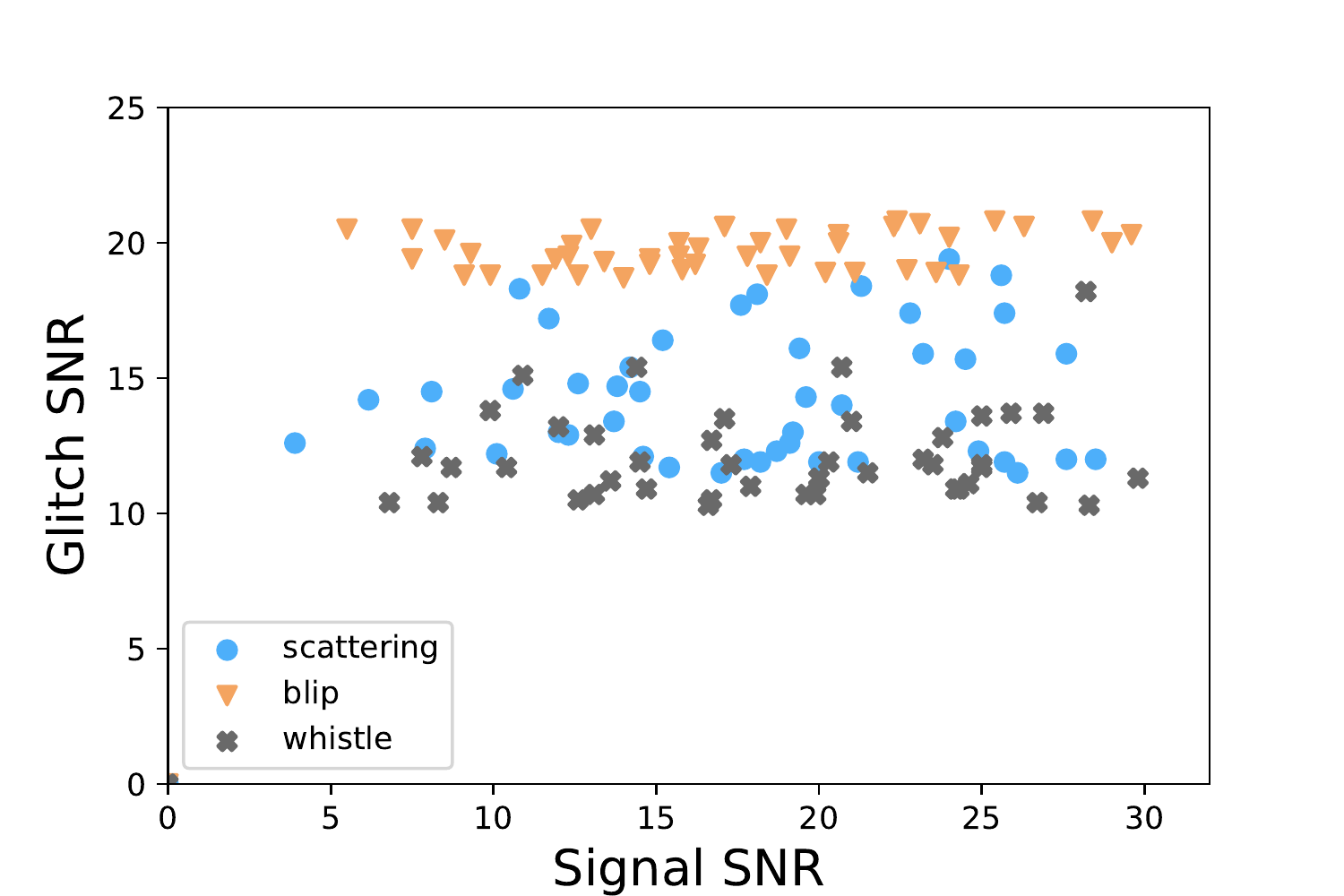}
\caption{The log Bayes factors for BBH signals when a glitch is present. The line shows the average expected value when no glitches are present. (Top left) The glitch is directly on top of the signal. (Top right) The glitch is 0.1\,s away from the signal. (Bottom left) The glitch is 0.2\,s away from the signal. (Bottom right) The SNR values for each BBH signal and glitch pair. Blip glitches create the largest error in the measured log Bayes factors.}
\label{fig:bbh_logb}
\end{figure}

As the log Bayes factors can be used as a detection statistic and are important for model selection, first we calculate the log Bayes factors for all of the signals at all time offsets from the glitches. The results are shown in Figure \ref{fig:bbh_logb}. 

When the signal in the data matches the signal model, the log Bayes factors are expected to be proportional to the square of the SNR values. When the BBH signal is exactly on top of the glitch, the glitch that has the biggest effect on the results is the blip glitches. The blip glitches increase the value of the log Bayes factors, which could artificially increase the confidence in a detection, or make the signal model look more probable than it should be. A very small number $(\sim 3)$ of the scattered light glitches resulted in a small increase in the log Bayes factors. The whistle glitches do not have any effect on the log Bayes factors. When the glitch is 0.1\,s away from the signal, the blip glitches still have a large effect on the log Bayes factors, but the number of glitches that increase the log Bayes factors is around half the number that effected the signals when the signal was directly on top of the glitch. When the glitch is 0.2\,s away from the signal, none of the glitches have a strong effect on the the log Bayes factors. 

\begin{figure}[!t]
\centering
\includegraphics[width=0.48\textwidth,height=4.5cm]{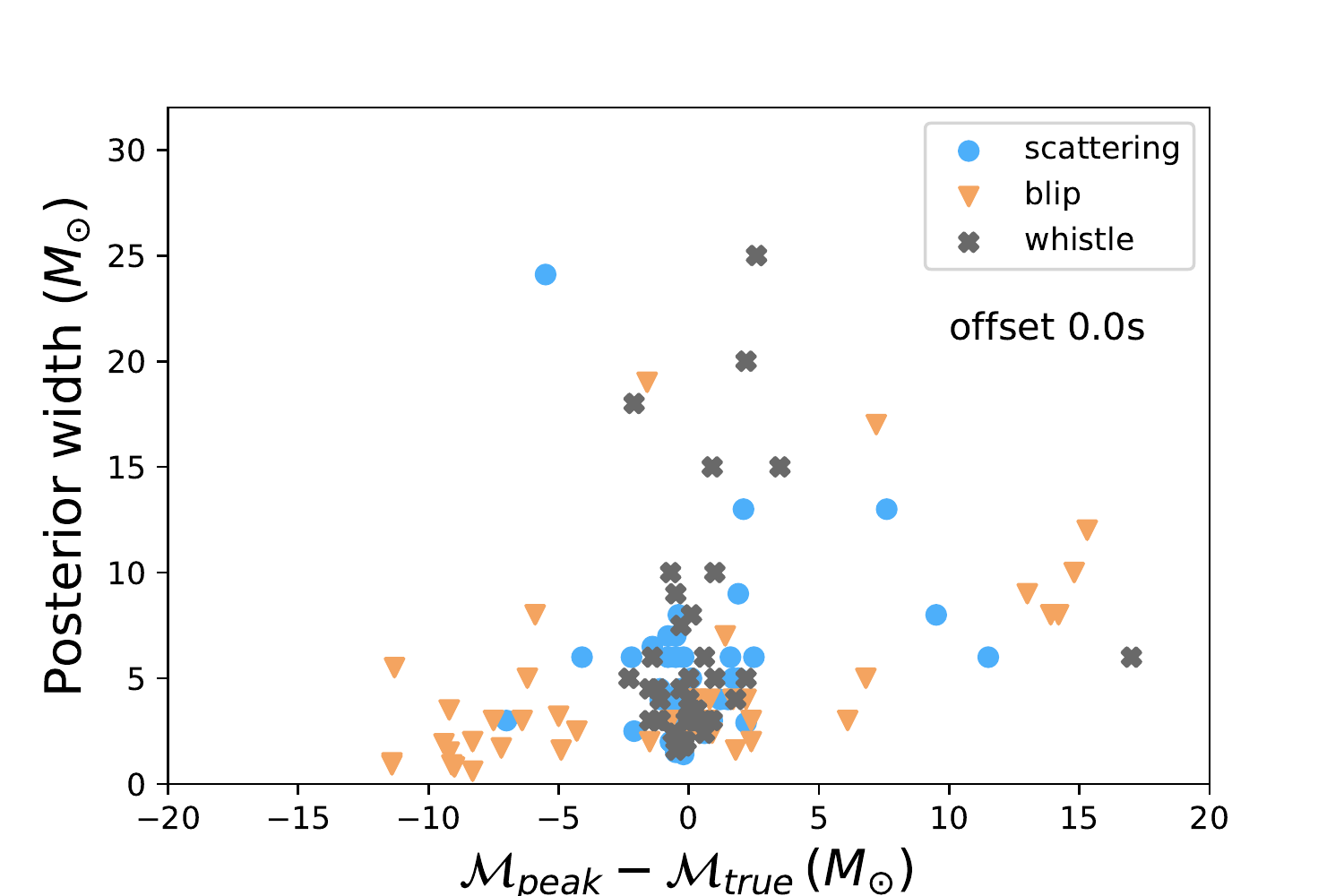}
\includegraphics[width=0.48\textwidth,height=4.5cm]{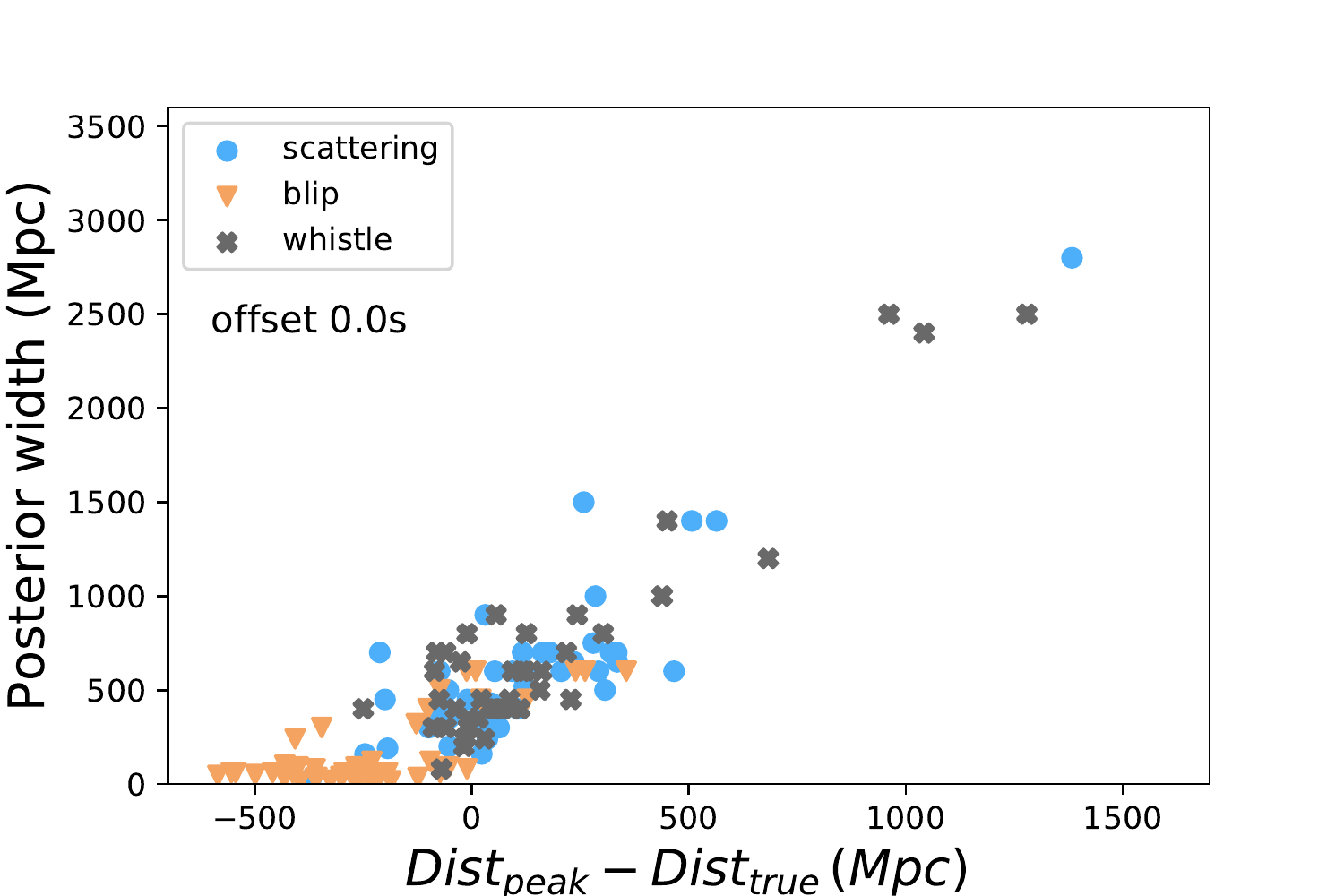}
\includegraphics[width=0.48\textwidth,height=4.5cm]{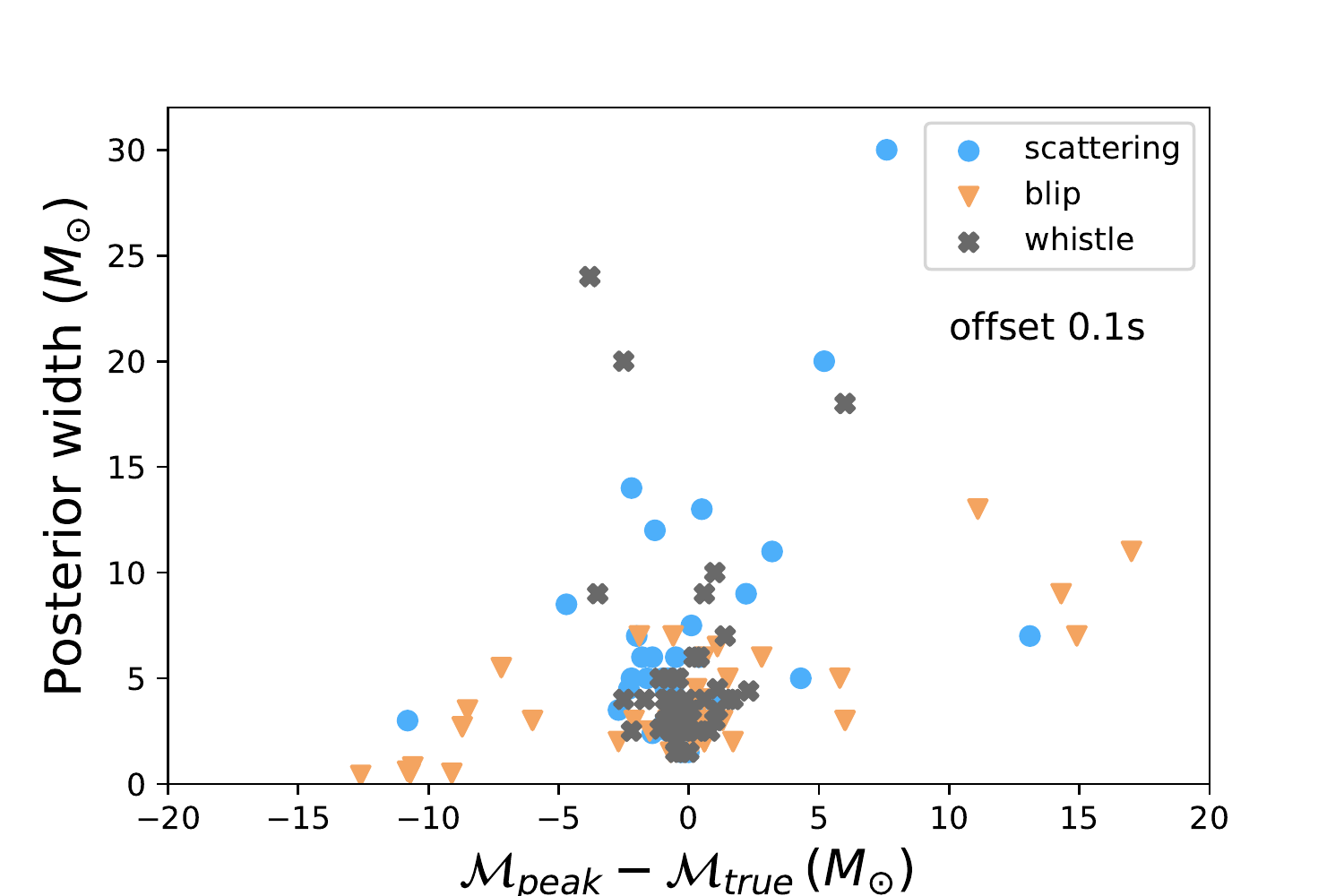}
\includegraphics[width=0.48\textwidth,height=4.5cm]{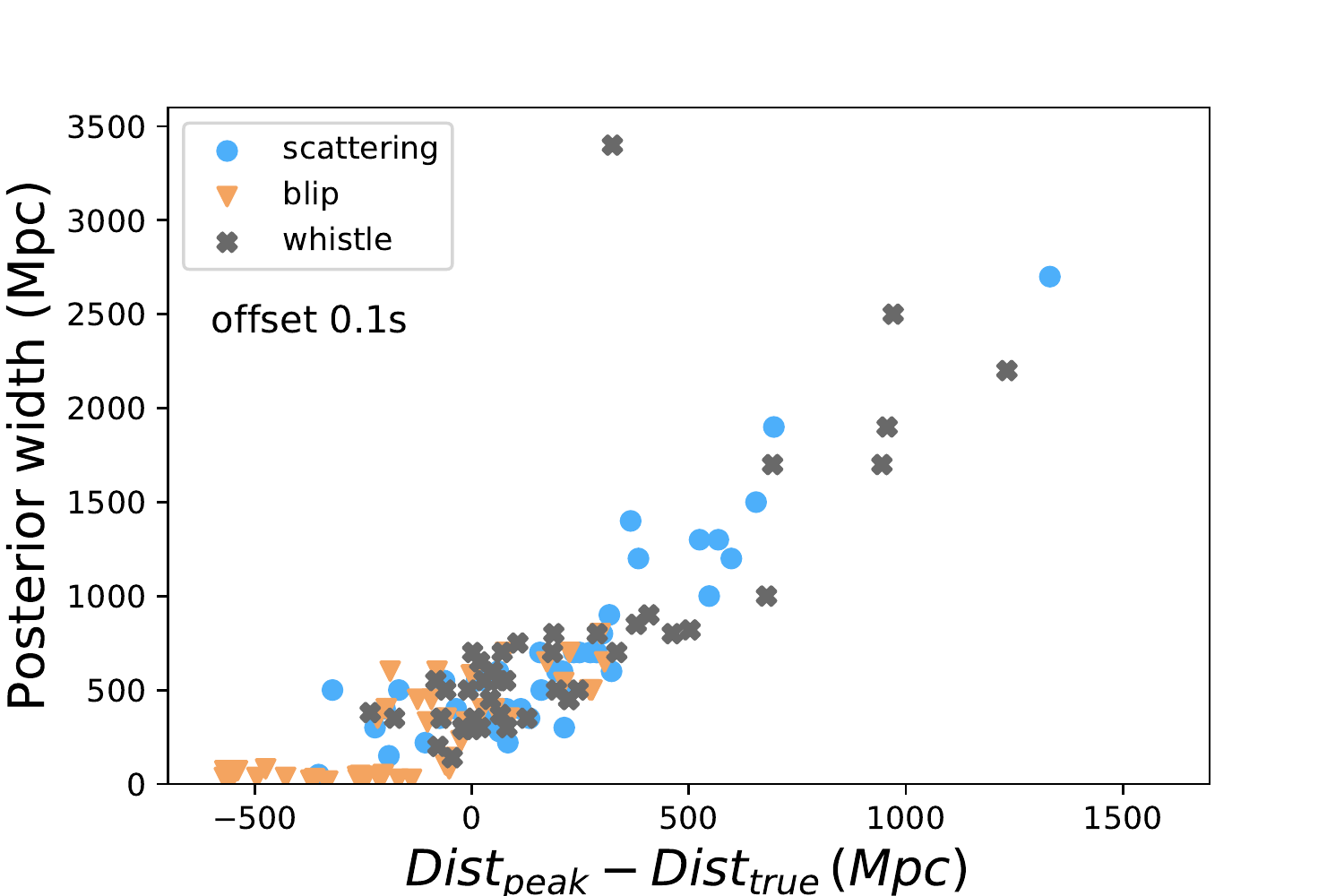}
\includegraphics[width=0.48\textwidth,height=4.5cm]{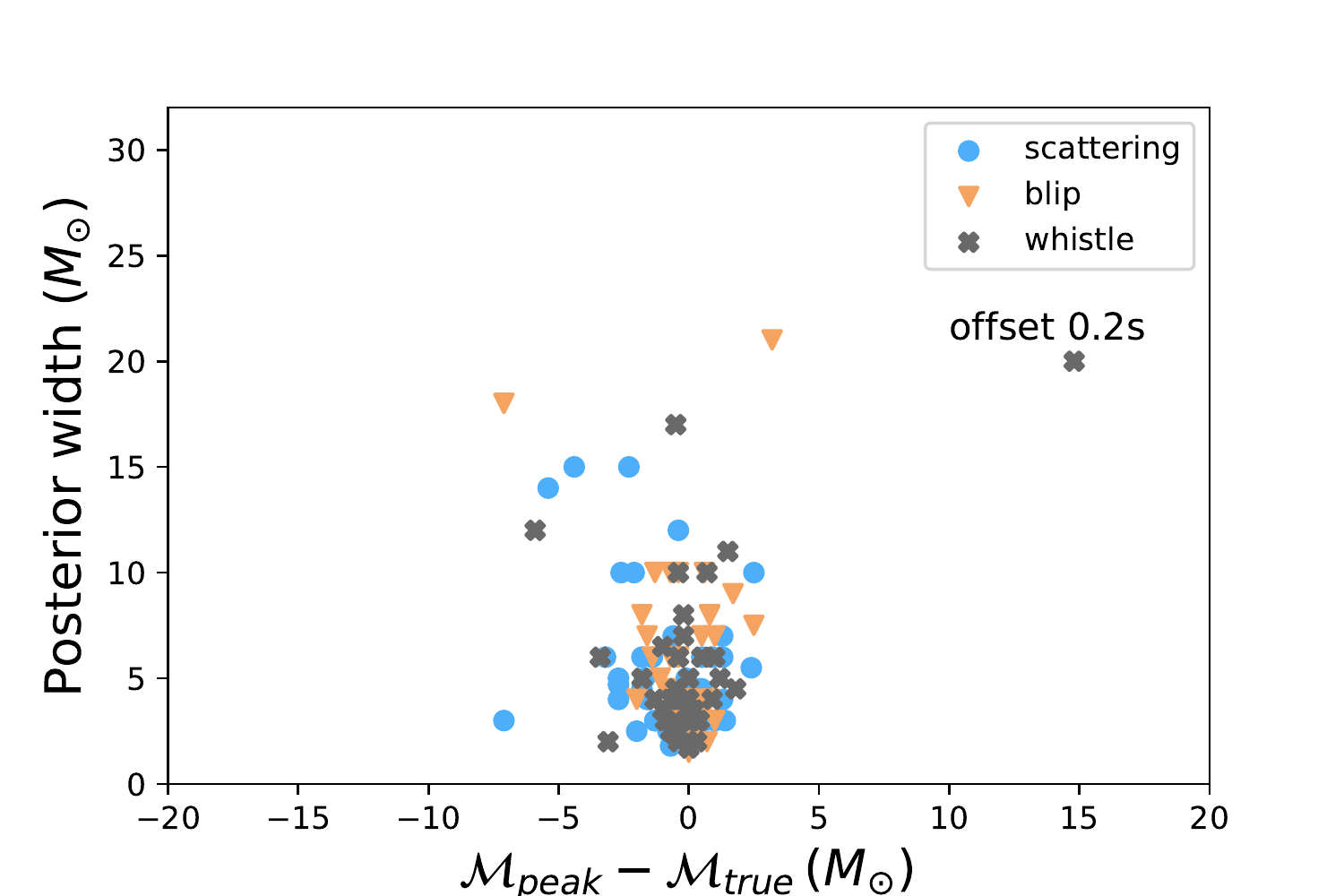}
\includegraphics[width=0.48\textwidth,height=4.5cm]{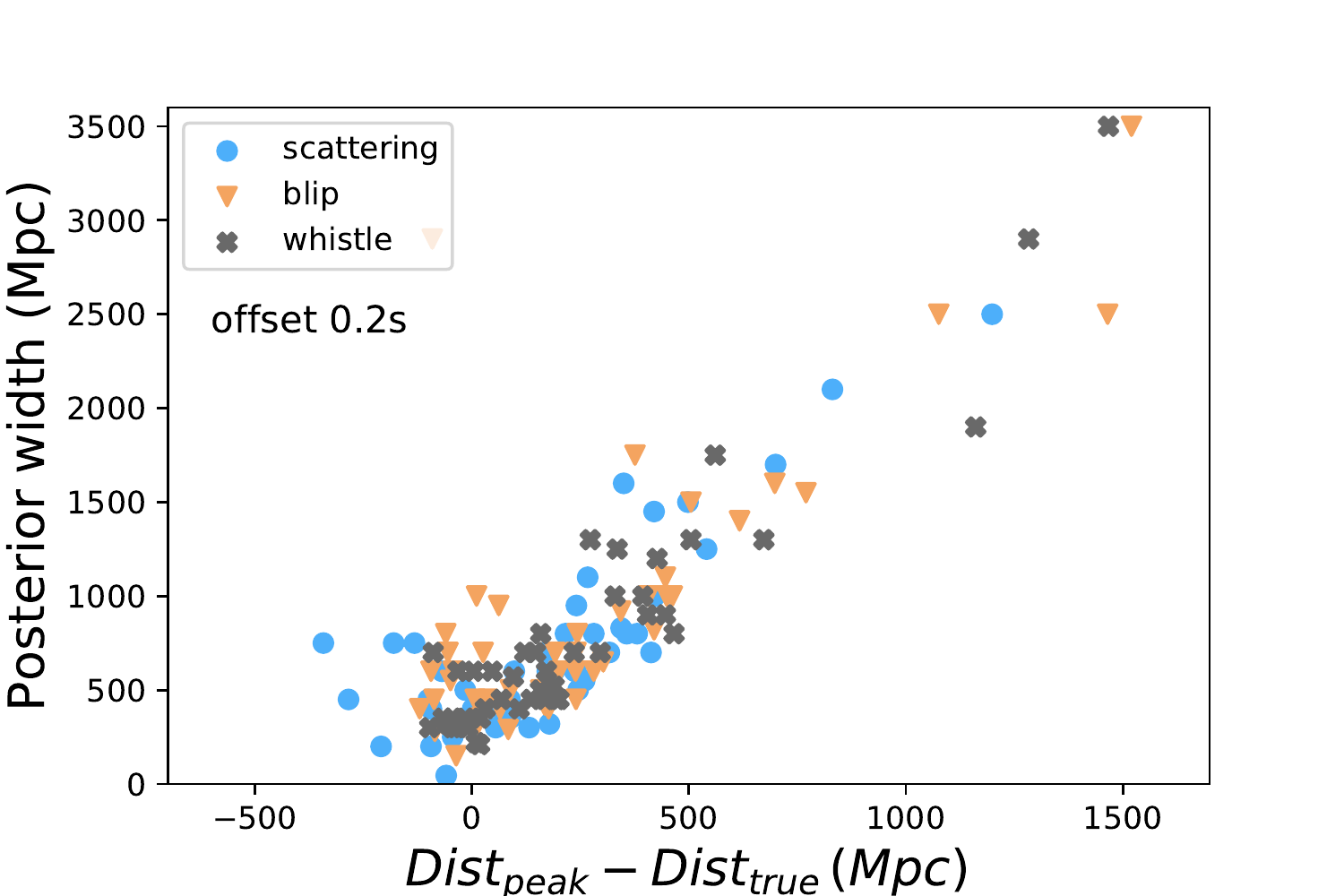}
\caption{The difference between the peak of the posteriors and the true chirp mass and distance parameters and the width of the 90\% confidence intervals of the posterior distributions. (Top) The glitch is directly on top of the signal. (Middle) The glitch is 0.1\,s away from the signal. (Bottom) The glitch is 0.2\,s away from the signal. The blip glitches create the biggest errors on both the chirp mass and distance posterior distributions. }
\label{fig:bbh_conf}
\end{figure} 

The parameter estimation results for the BBH signals are summarized in Figure \ref{fig:bbh_conf}. We look at the difference between the peak of the posterior distributions and the true values, and the width of the 90\% confidence intervals of the posterior distributions for the distance and chirp mass parameters. When the signal is directly on top of the glitch, the blip glitches have the largest effect on the difference between the injected value and the posterior peak. A small number of scattered light and whistle glitches make the chirp mass posterior distributions wider, therefore increasing the error on the measured parameters. The blip glitches artificially increase the amplitudes of the signals making the distances appear to be smaller than the actual value, and they make the posterior widths much smaller than expected, which makes the error on the distance much smaller than it should be. The distance values have larger posterior widths at larger distances, as signals at larger distances have a smaller SNR.

When the signal is 0.1\,s away from the glitch, we see the same effects as when the signal is directly on top of the glitch,  but for only half as many glitches and signals as in the previous case. The blip glitches increase the difference between the peak of the posteriors and the true parameter values, and the longer duration whistle glitches and scattered light increase the errors on the chirp mass measurements. When the signal is 0.2\,s away from the glitch, the glitches no longer have a large effect on the measured chirp mass and distance. The glitch and BBH signal SNR values are shown in Figure \ref{fig:bbh_conf}. At all distances, the worst effects are found when the SNR of the glitch is larger than the SNR of the signals.

\subsection{Example BBH posteriors}

\begin{figure}[!t]
\centering
\includegraphics[width=0.49\textwidth,height=4.5cm]{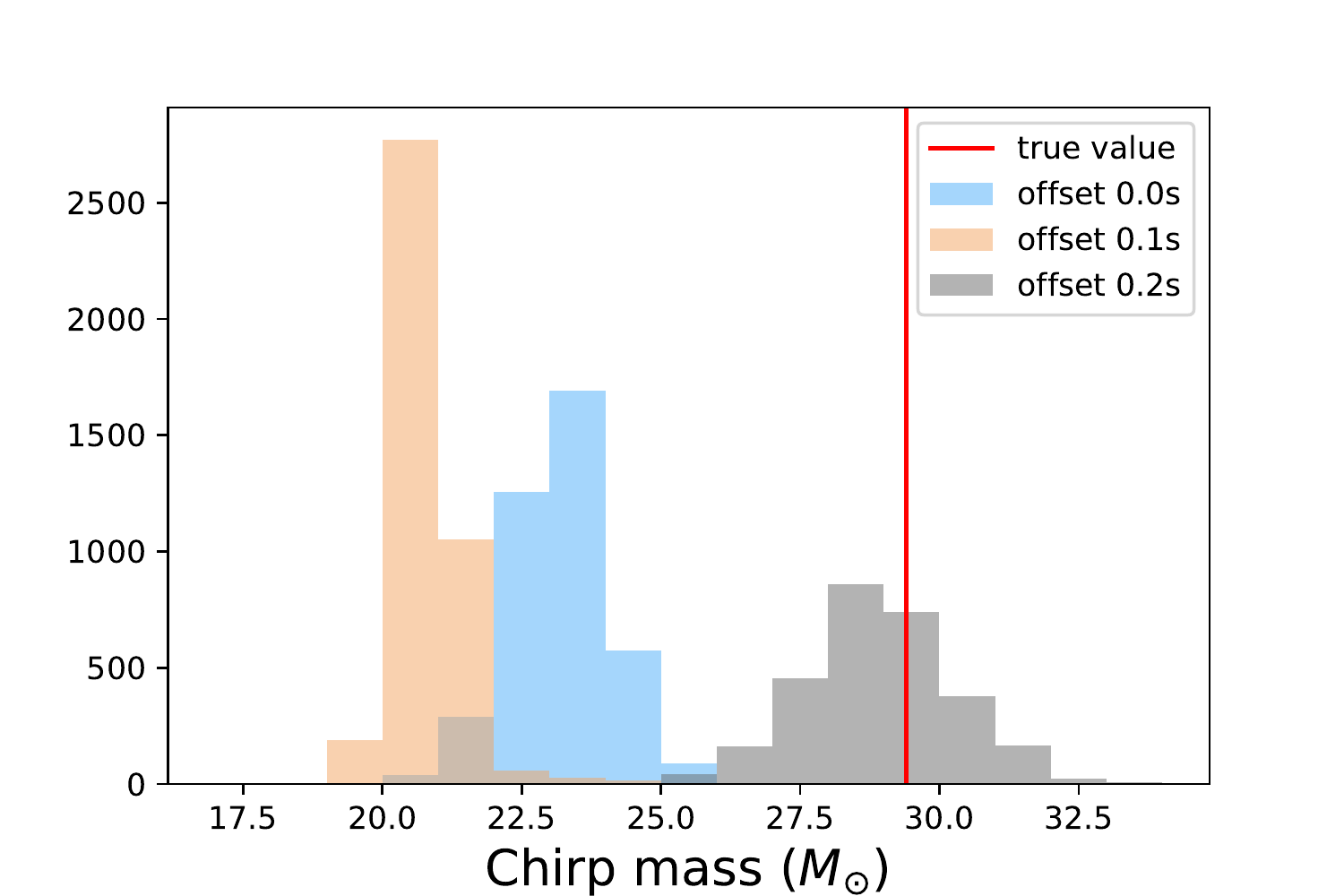}
\includegraphics[width=0.49\textwidth,height=4.5cm]{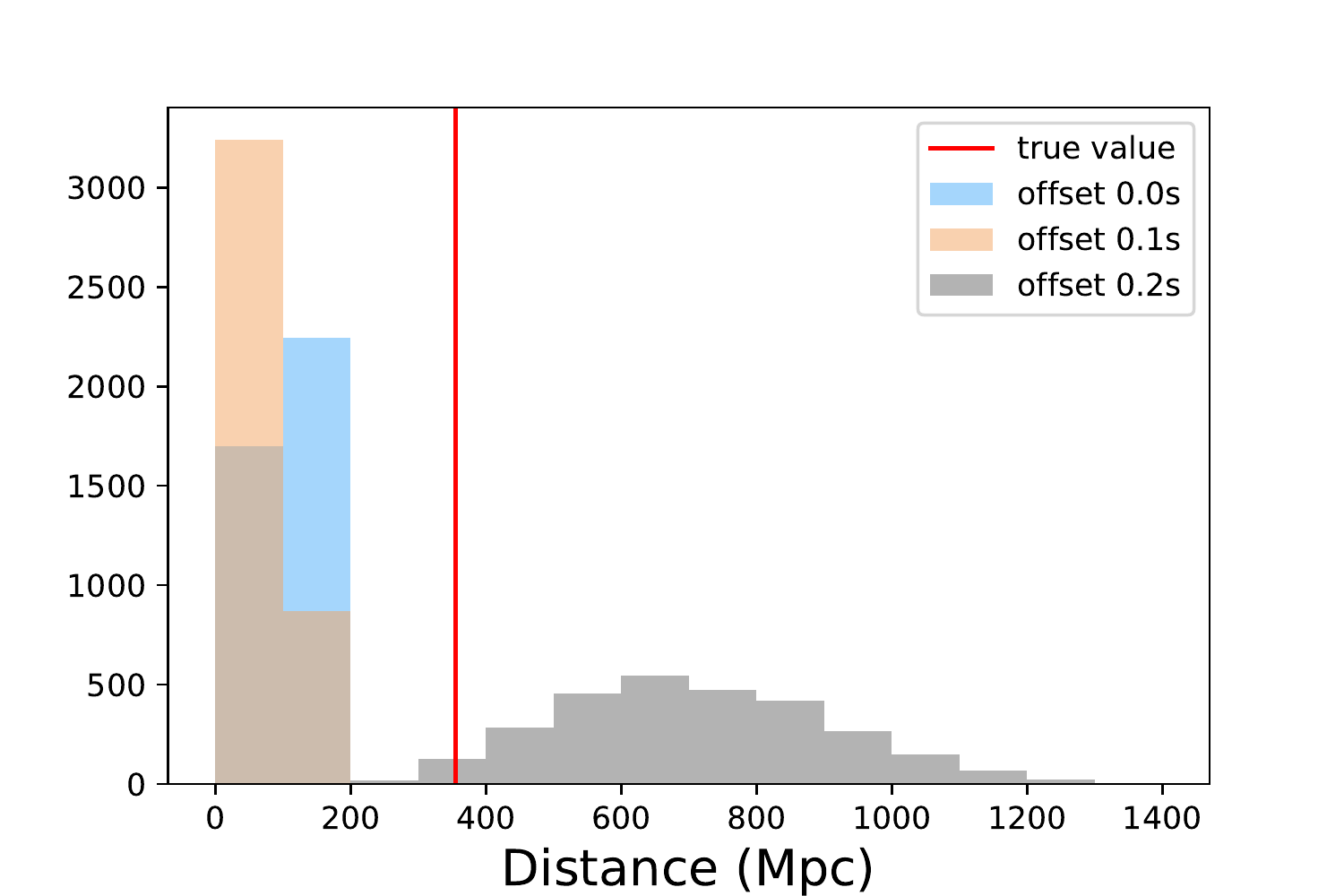}
\caption{An example of the chirp mass and distance posteriors for a BBH signal at increasing time offsets from scattered light. The values are smaller than expected when the glitch has an offset of 0.1\,s or less.}
\label{fig:bbh_posts}
\end{figure}

In this subsection, we show in more detail the results from one of the artificial BBH signals that was badly affected by the glitch. The signal has a chirp mass of $29.4\,M_{\odot}$, is at a distance of 355\,Mpc, and has a network SNR of 11.9. It is injected on top of a scattered light glitch with an SNR of 19.4. In Figure \ref{fig:bbh_posts}, we show the posteriors at increasing time offsets from the glitch. When the signal is directly on top of the glitch, or 0.1\,s seconds away, the injected values are outside of the posterior distributions. The extra amplitude provided by the glitch, and the frequency of the glitch, makes the signal appear to have a smaller distance and chirp mass. The injected values are inside the posterior distributions when the signal is 0.2\,s away from the glitch.  

\section{Sine Gaussian burst results}
\label{sec:results_sg}

\begin{figure}[!t]
\centering
\includegraphics[width=0.49\textwidth,height=4.5cm]{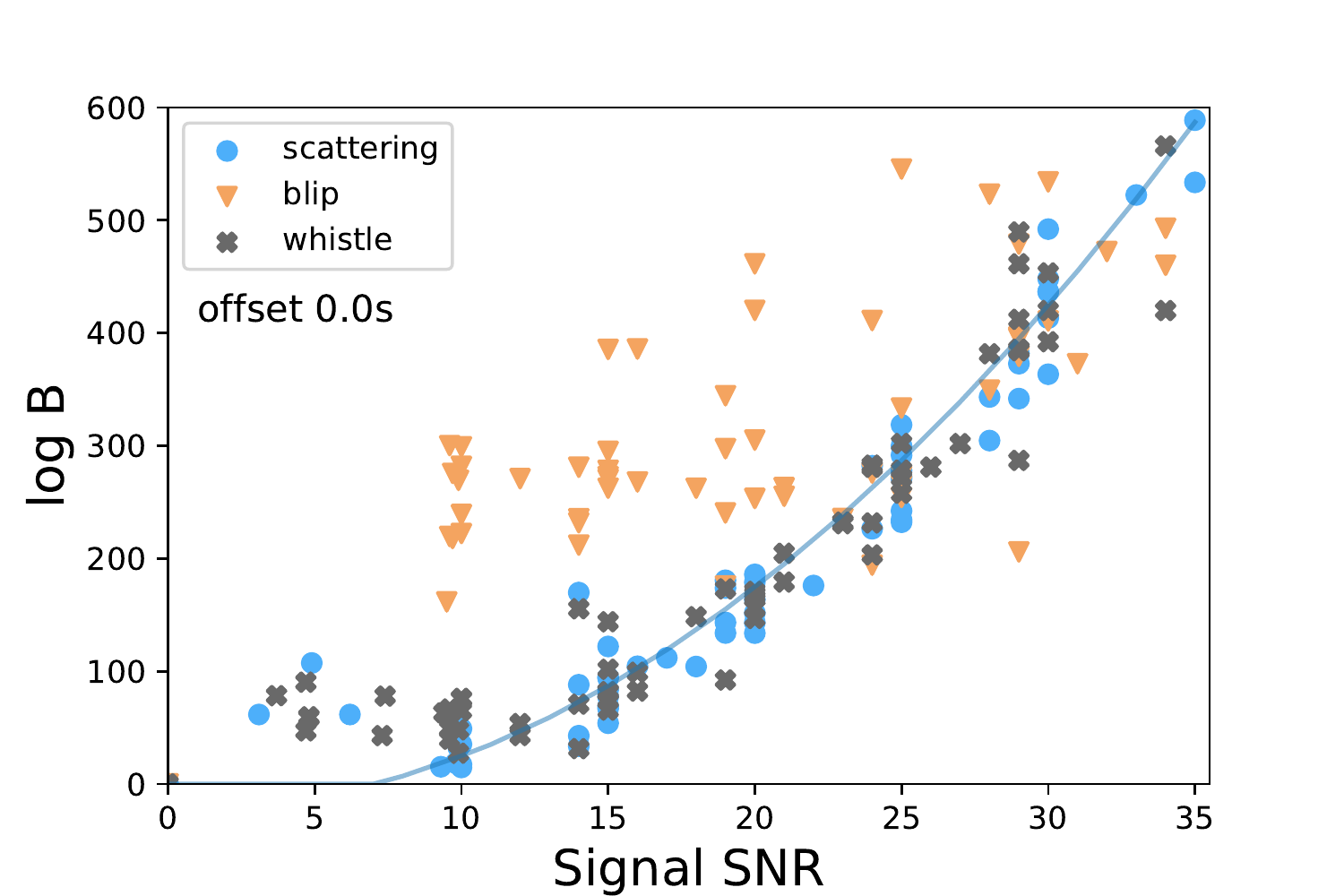}
\includegraphics[width=0.49\textwidth,height=4.5cm]{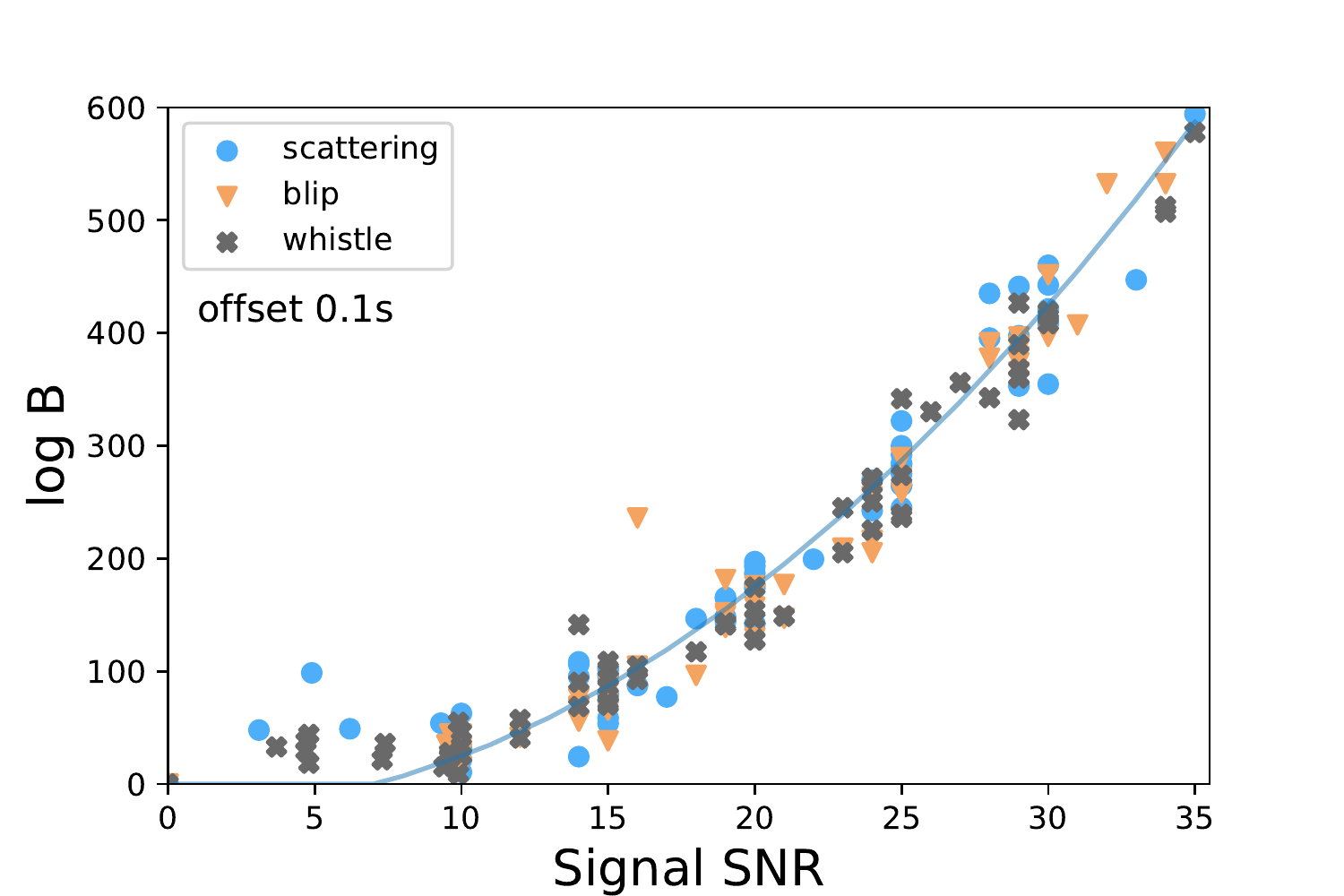}
\includegraphics[width=0.49\textwidth,height=4.5cm]{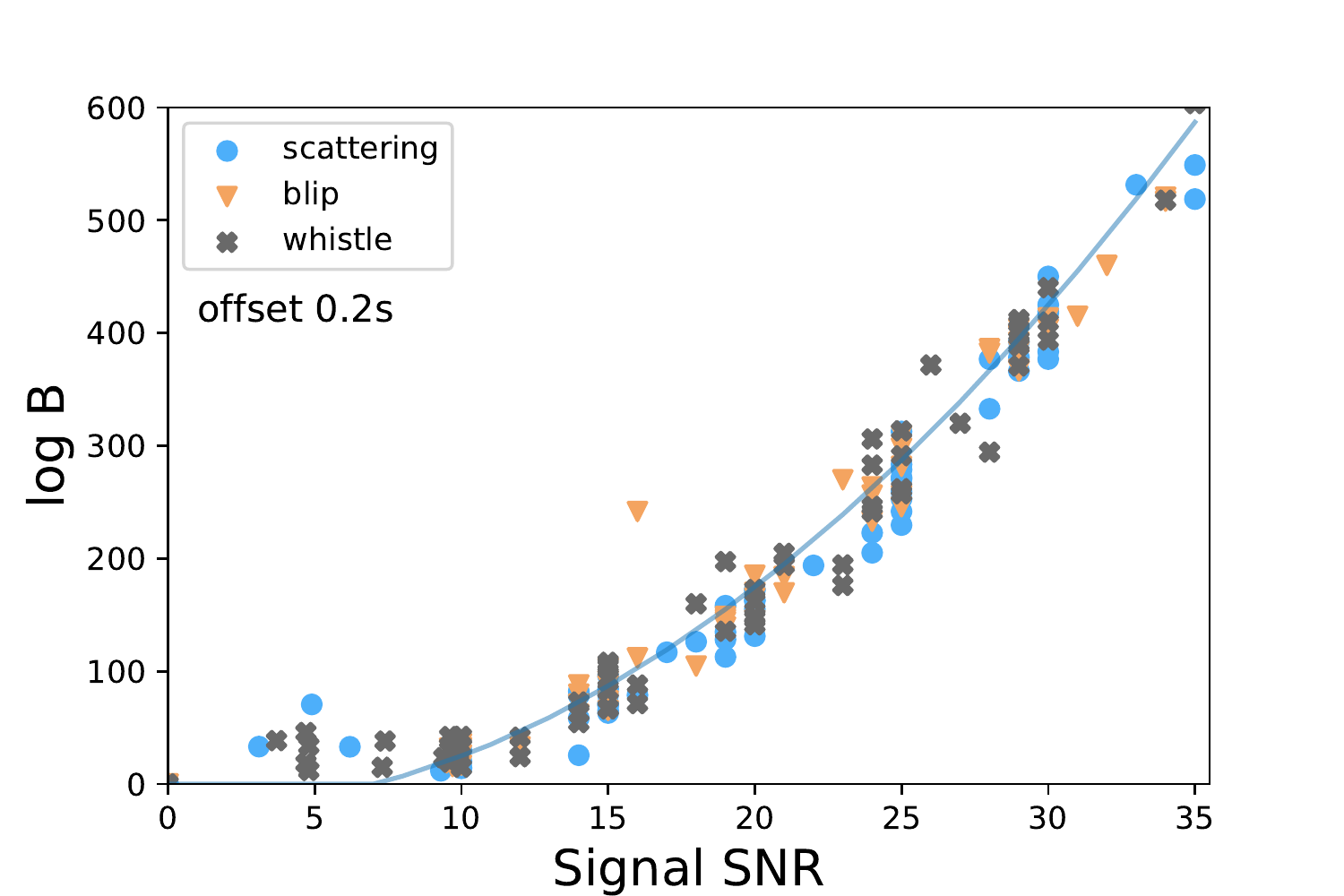}
\includegraphics[width=0.49\textwidth,height=4.5cm]{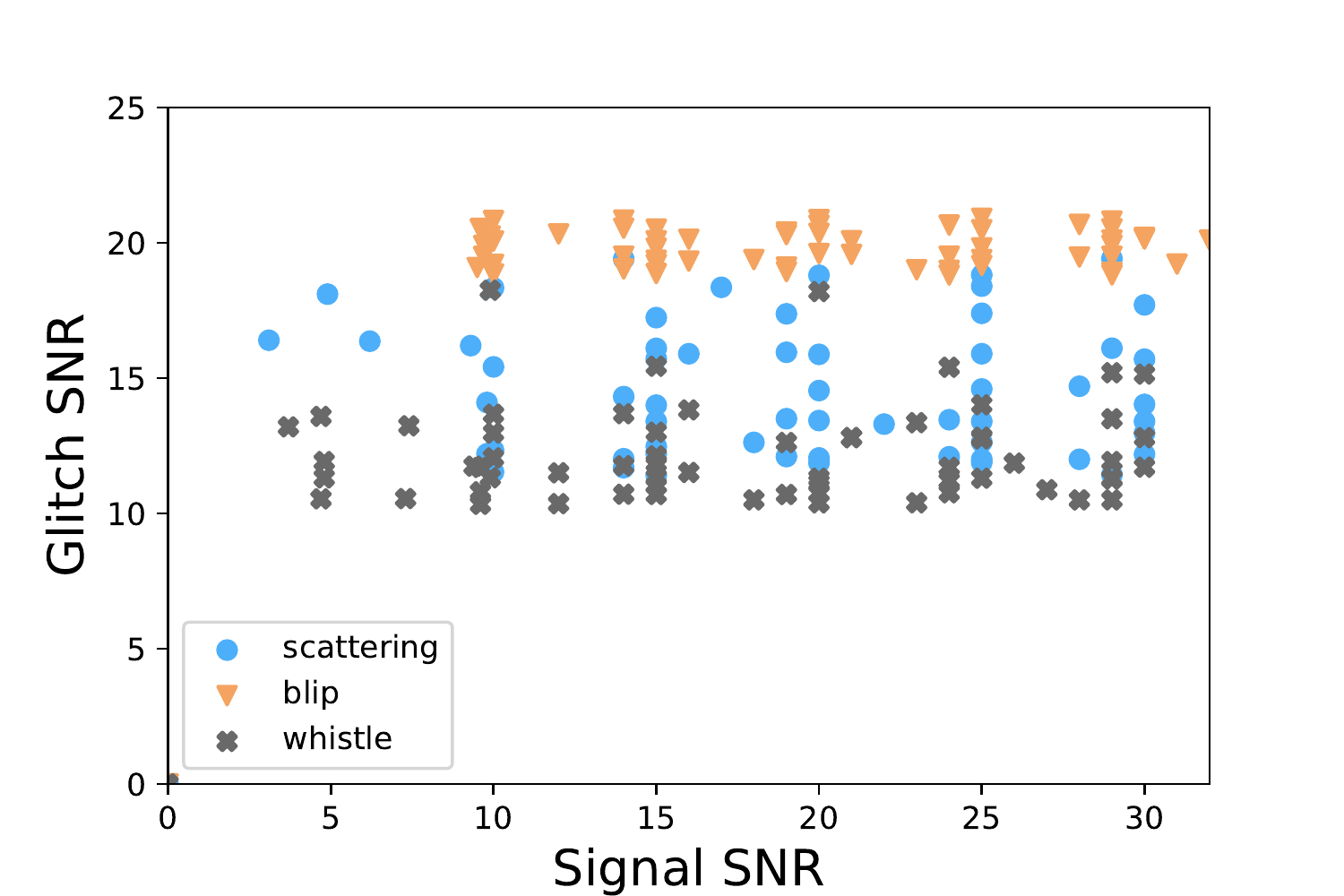}
\caption{The log Bayes factors for sine Gaussian signals when a glitch is present. The line shows the average expected value when no glitches are present. (Top left) The glitch is directly on top of the signal. (Top right) The glitch is 0.1\,s away from the signal. (Bottom left) The glitch is 0.2\,s away from the signal. (Bottom right) The SNR values for each sine Gaussian signal and glitch pair. The blip glitches increase the log Bayes factors for signals with an SNR larger than 10, and the whistle and scattered light glitches increase the log Bayes factors of signals with an SNR less than 10.}
\label{fig:sg_logb}
\end{figure}

In this section, we describe the results for sine Gaussian burst signals injected on top of glitches and recovered with a sine Gaussian signal model. The log Bayes factors, at all time offsets, are shown in Figure \ref{fig:sg_logb}. As for the BBH results, when the signal is injected directly on top of the glitch the blip glitches create a large increase in the log Bayes factors for the signals with an SNR larger than 10. When the signal network SNR is less than 10, there is an increase in the Bayes factors of the signals injected on top of the whistle and scattered light glitches. When the signal is 0.1\,s and 0.2\,s away from the glitch, only one blip glitch creates an increase in the log Bayes factor. As for the BBH signals, the glitches have a greater effect on the signal if the glitch SNR is larger than the signal SNR. 

\begin{figure}[!t]
\centering
\includegraphics[width=0.48\textwidth,height=4.5cm]{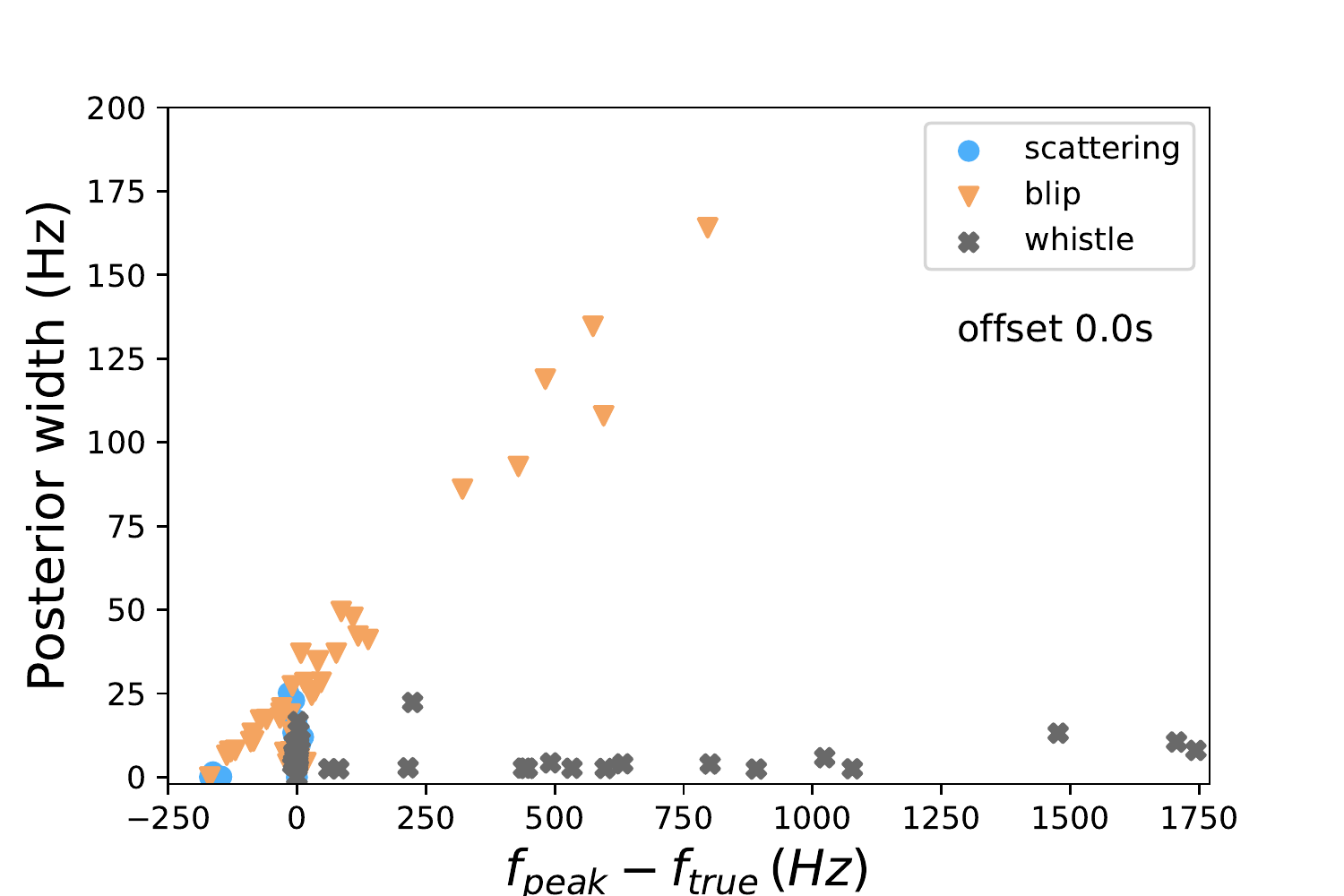}
\includegraphics[width=0.48\textwidth,height=4.5cm]{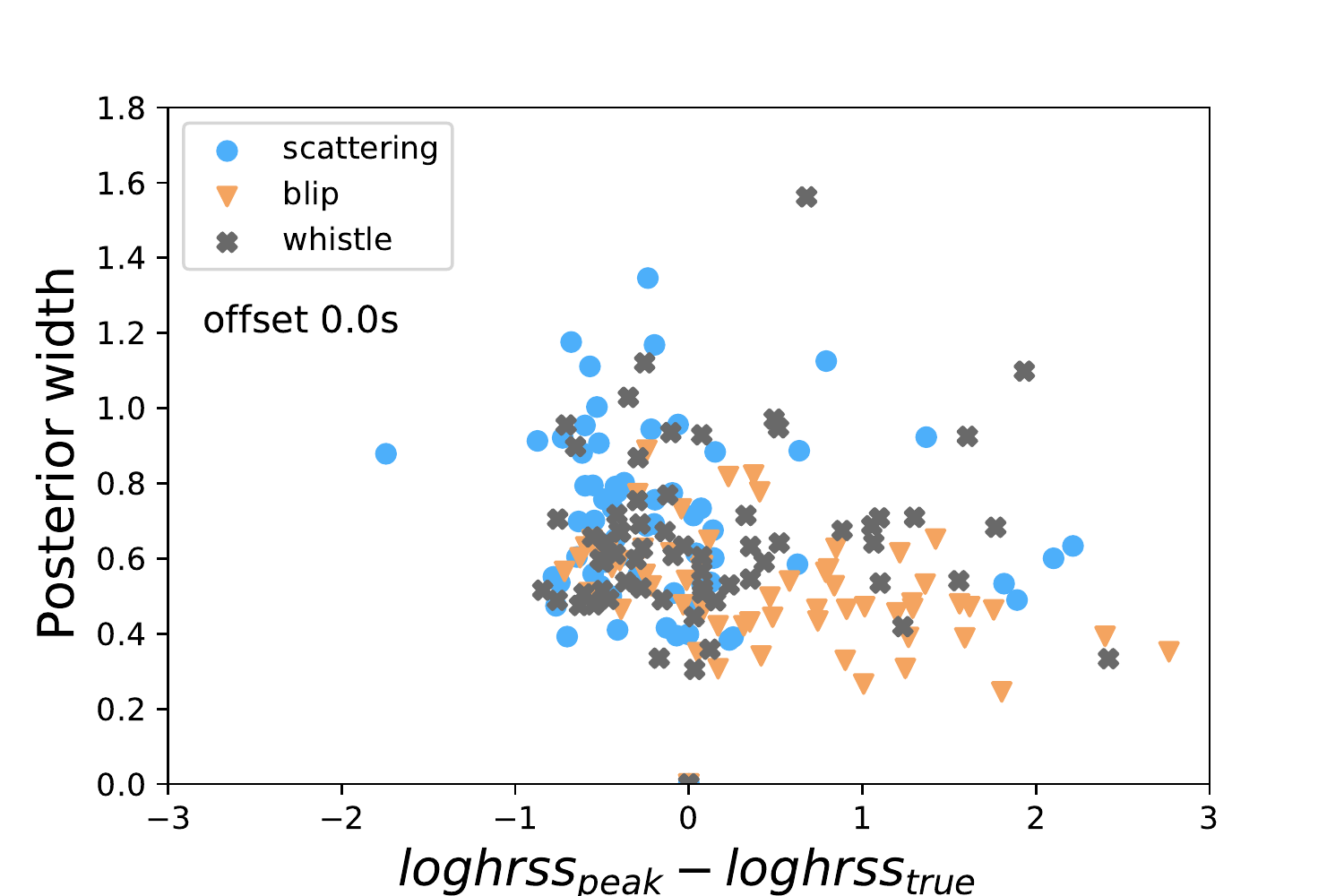}
\includegraphics[width=0.48\textwidth,height=4.5cm]{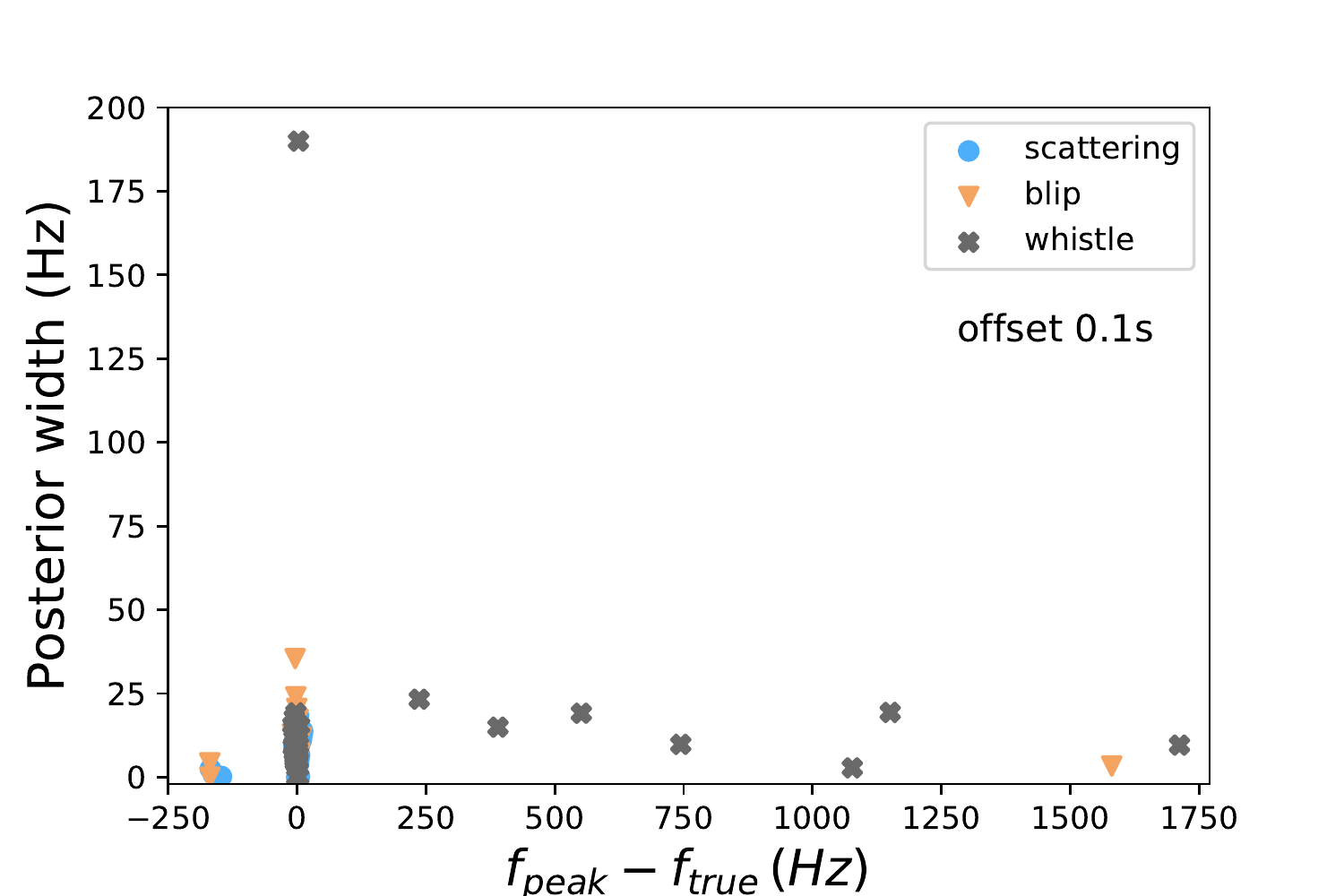}
\includegraphics[width=0.48\textwidth,height=4.5cm]{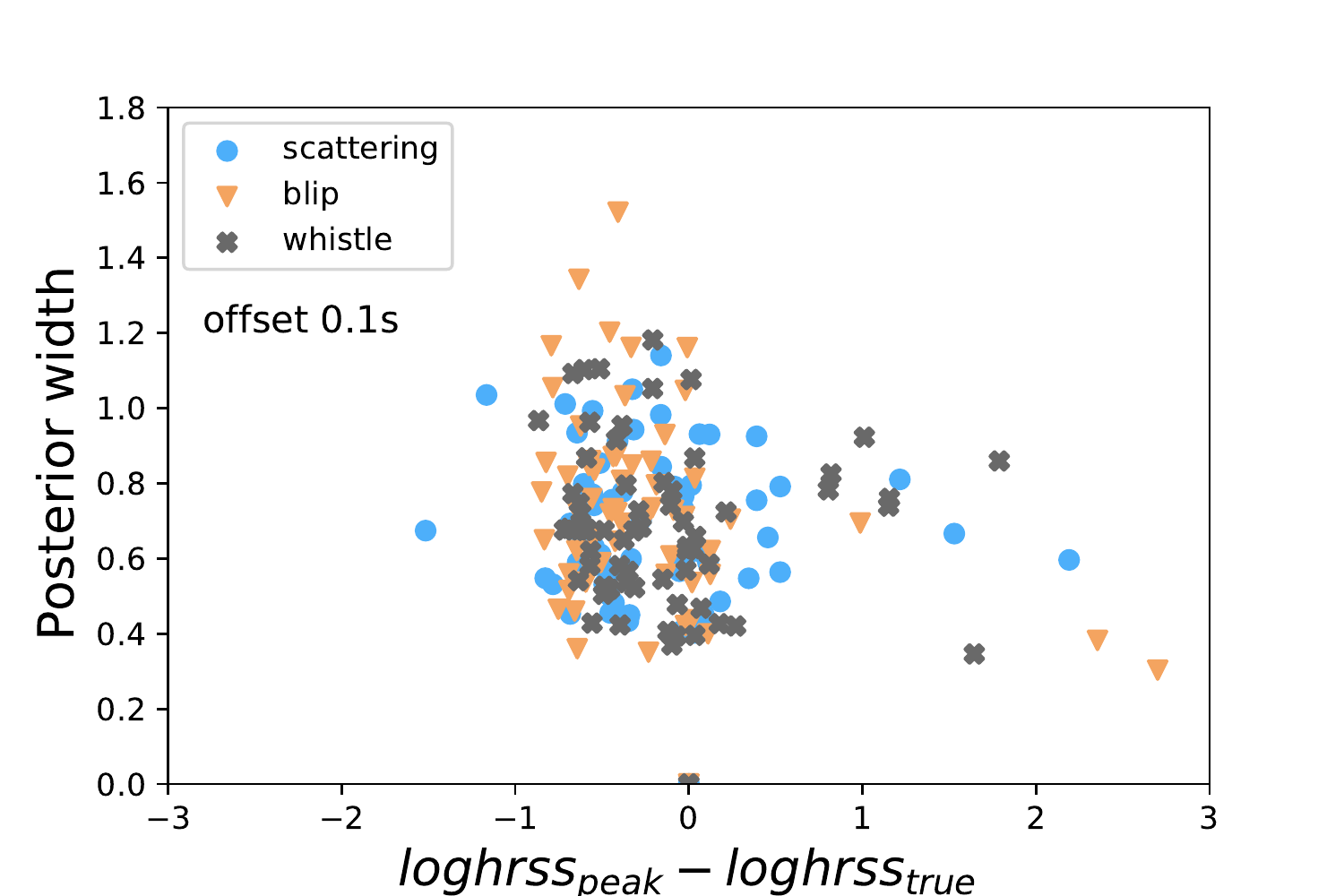}
\includegraphics[width=0.48\textwidth,height=4.5cm]{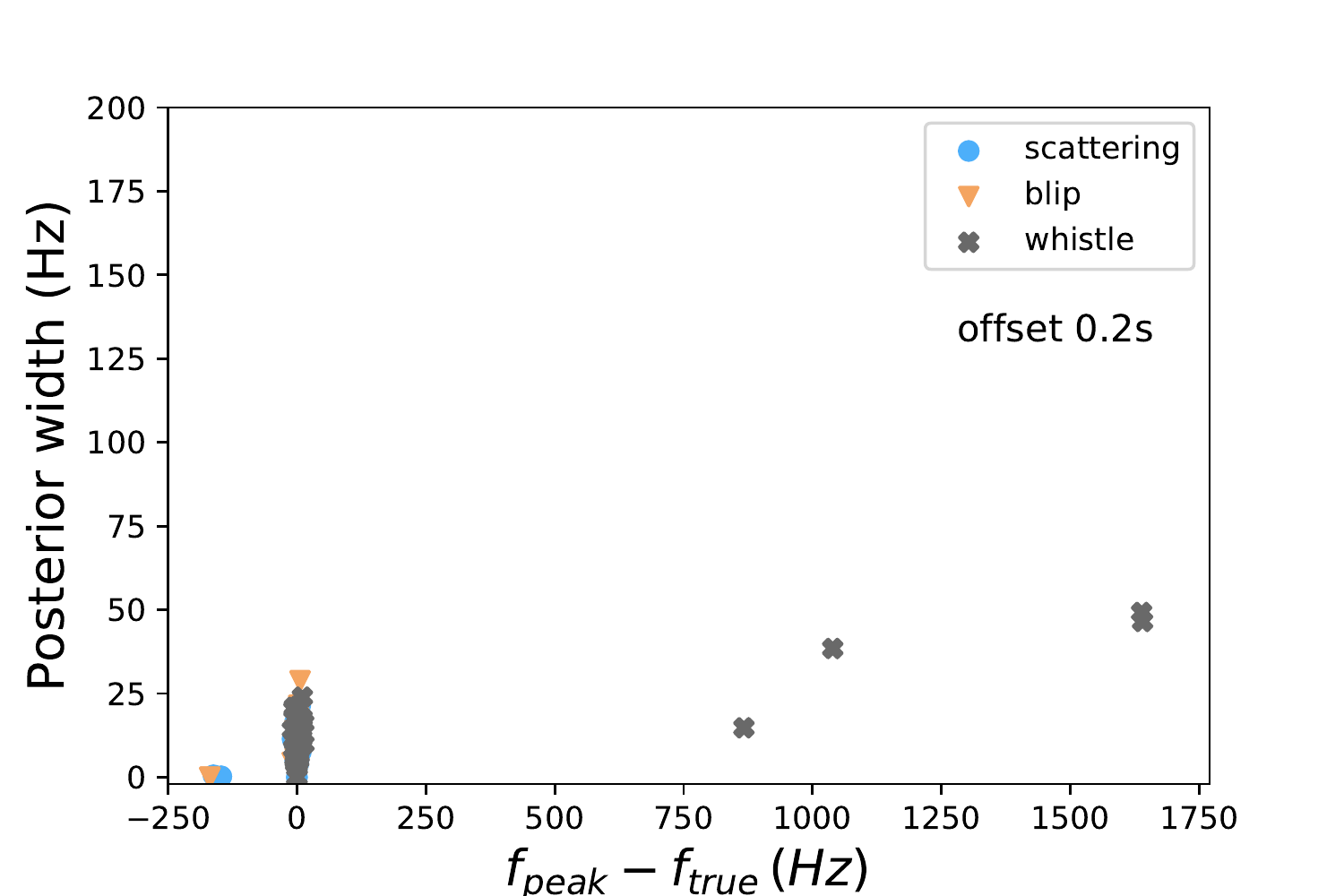}
\includegraphics[width=0.48\textwidth,height=4.5cm]{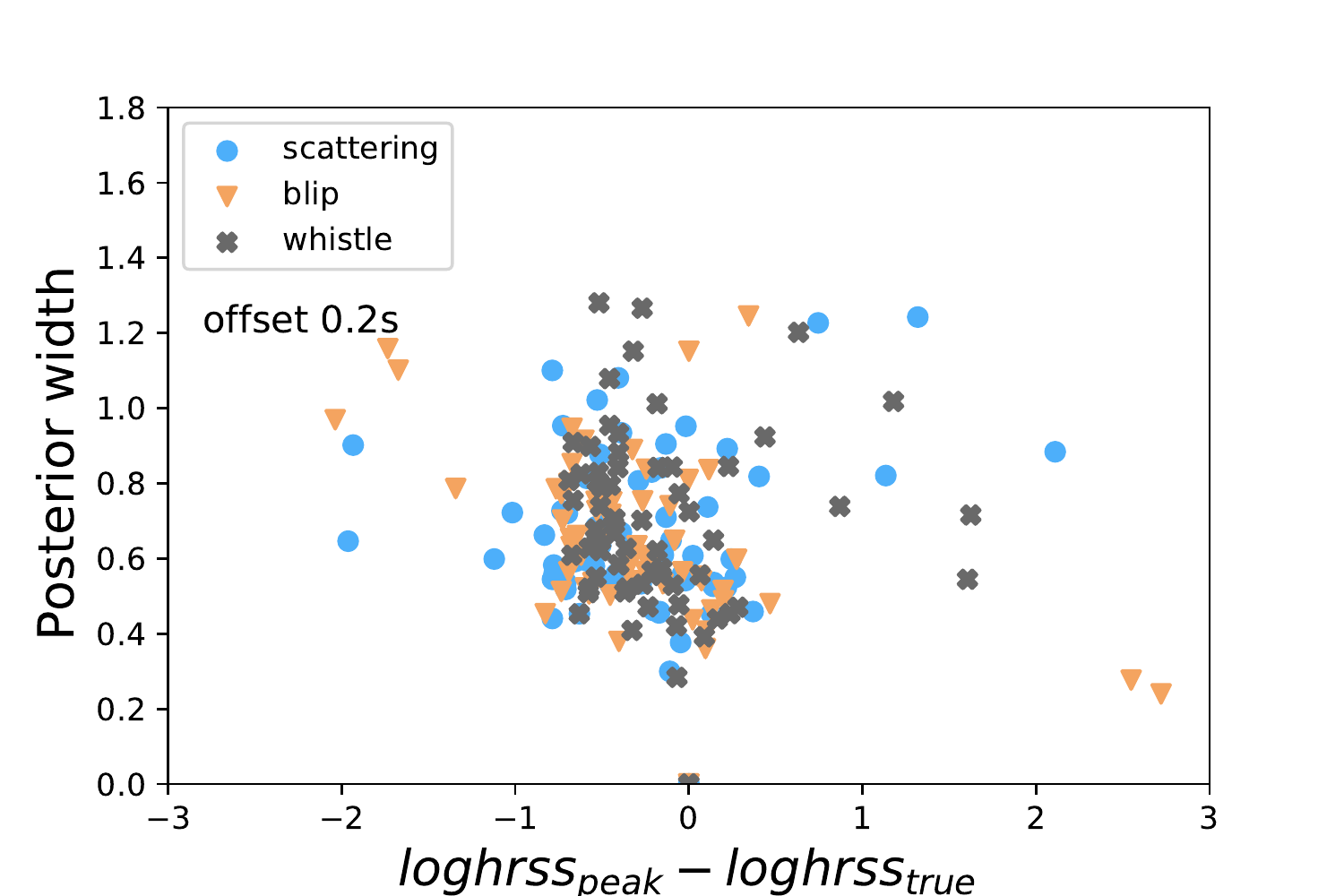}
\caption{The difference between the peak of the posteriors and the true values for the frequency and hrss parameters of the sine Gaussian signals, and the width of the 90\% confidence intervals of the posterior distributions. (Top) The glitch is directly on top of the signal. (Middle) The glitch is 0.1\,s away from the signal. (Bottom) The glitch is 0.2\,s away from the signal. The whistle glitches make the frequency appear much higher than the true value.}
\label{fig:sg_conf}
\end{figure}

The parameter estimation results for the sine Gaussian signals recovered with a sine Gaussian signal model are shown in Figure \ref{fig:sg_conf}. When the signal is injected directly on top of the glitch, we find that the whistle glitches have a large effect on the peak of the frequency posteriors. As they are much higher frequency than the signals, the posterior peaks are much higher than the injected values. The blip glitches effected both the posterior peak frequency value and the posterior width, making the error on the frequency parameter larger. The blip glitches make the $\log(\mathrm{h}_{rss})$ values larger than the injected values. 

When the signal is 0.1\,s away from the glitch, the blip glitches no longer have any effect on the measured parameters. One whistle glitch creates a large increase in the width of the posterior, and 7 whistle glitches make the posterior peak frequency value much larger than the true value. When the signal is 0.2\,s away from the glitch, 4 of the whistle glitches make the frequency values higher than the true value. None of the glitches had a large effect on the $\log(\mathrm{h}_{rss})$ values when the signal was 0.1\,s or 0.2\,s away from the glitch.

\subsection{Example sine Gaussian posteriors}

\begin{figure}[!t]
\centering
\includegraphics[width=0.49\textwidth,height=4.5cm]{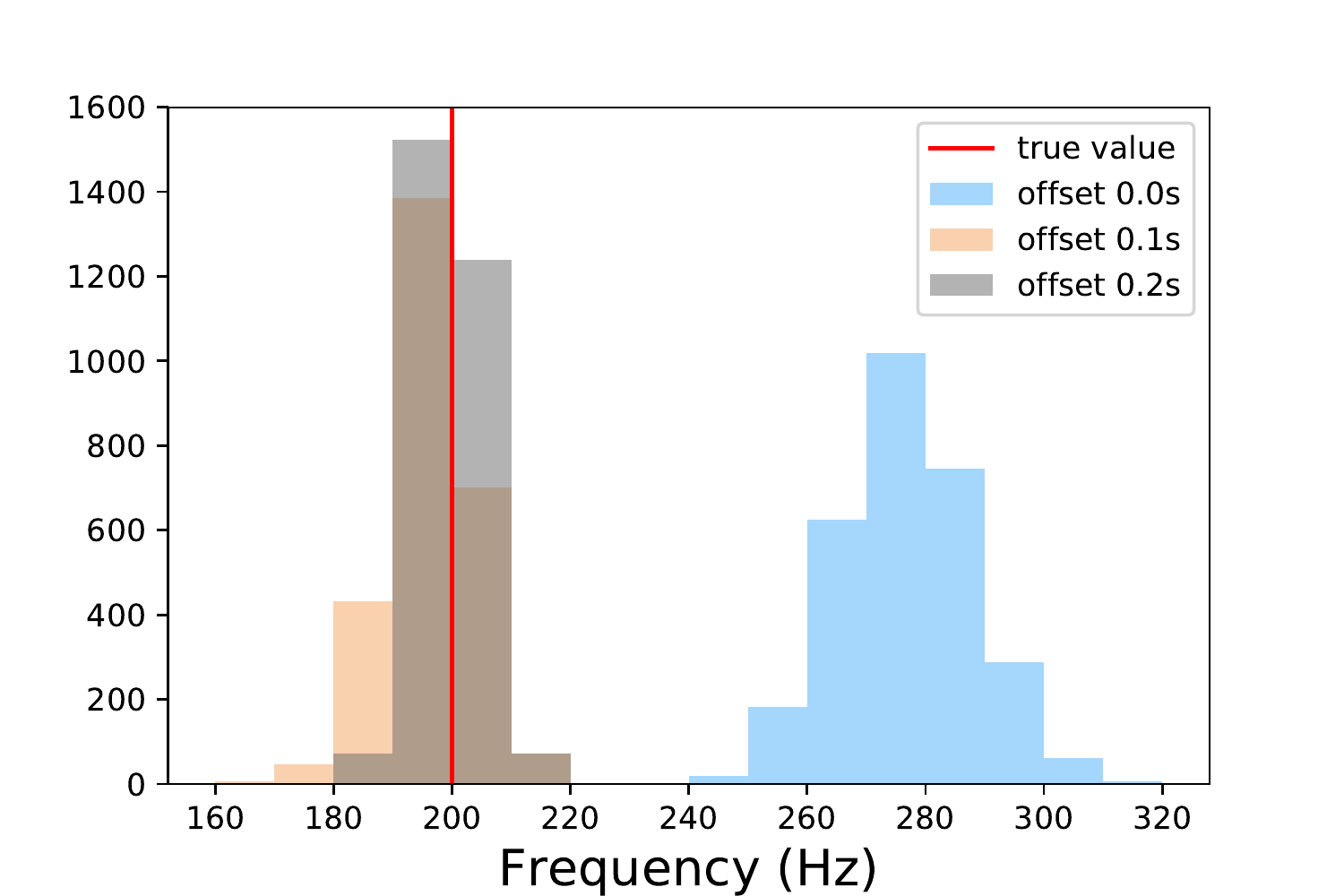}
\includegraphics[width=0.49\textwidth,height=4.5cm]{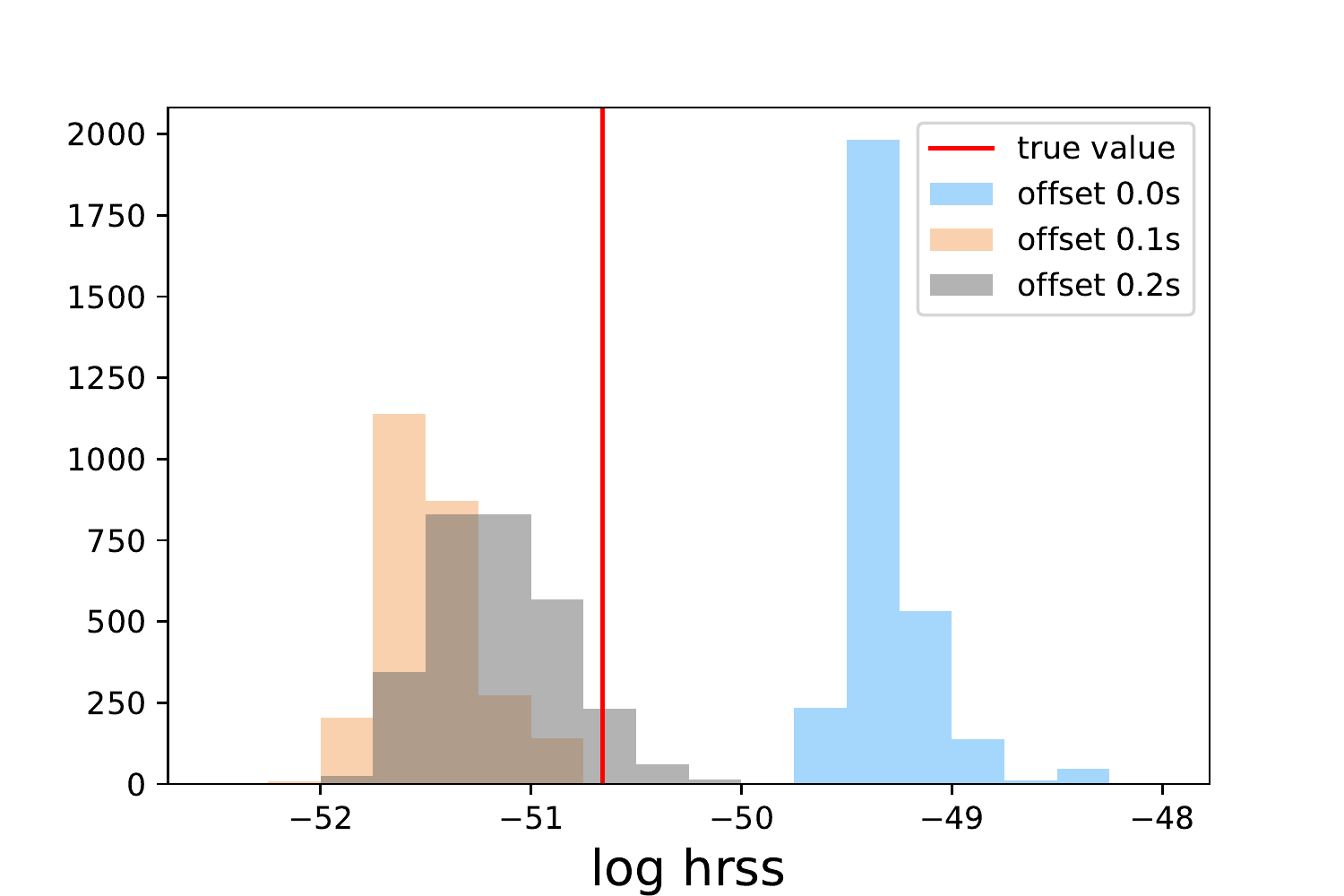}
\caption{An example of the frequency and log hrss posteriors for a sine Gaussian signal at increasing time offsets from a scattered light glitch. When the signal is directly on top of the glitch, the frequency and amplitude of the signal is larger than the true value. }
\label{fig:sg_posts}
\end{figure}

In this subsection, we show in more details the results of one of the sine Gaussian signals that was badly affected by a glitch. The signal has a frequency of 200\,Hz and has a network SNR of 9.7. It is injected on top of a scattered light glitch with an SNR of 19.9. In Figure \ref{fig:sg_posts}, we show the posteriors at increasing time offsets from the glitch. The posterior distributions show that the frequency and hrss values are larger than the true value when the signal is directly on top of the glitch. The true values are contained within the posteriors when the signal is 0.1\,s away from the glitch. The posterior distribution is narrower when the signal is 0.2\,s away from the glitch.  

\section{Supernova results}
\label{sec:results_sn} 

\begin{figure}[!t]
\centering
\includegraphics[width=0.49\textwidth,height=4.5cm]{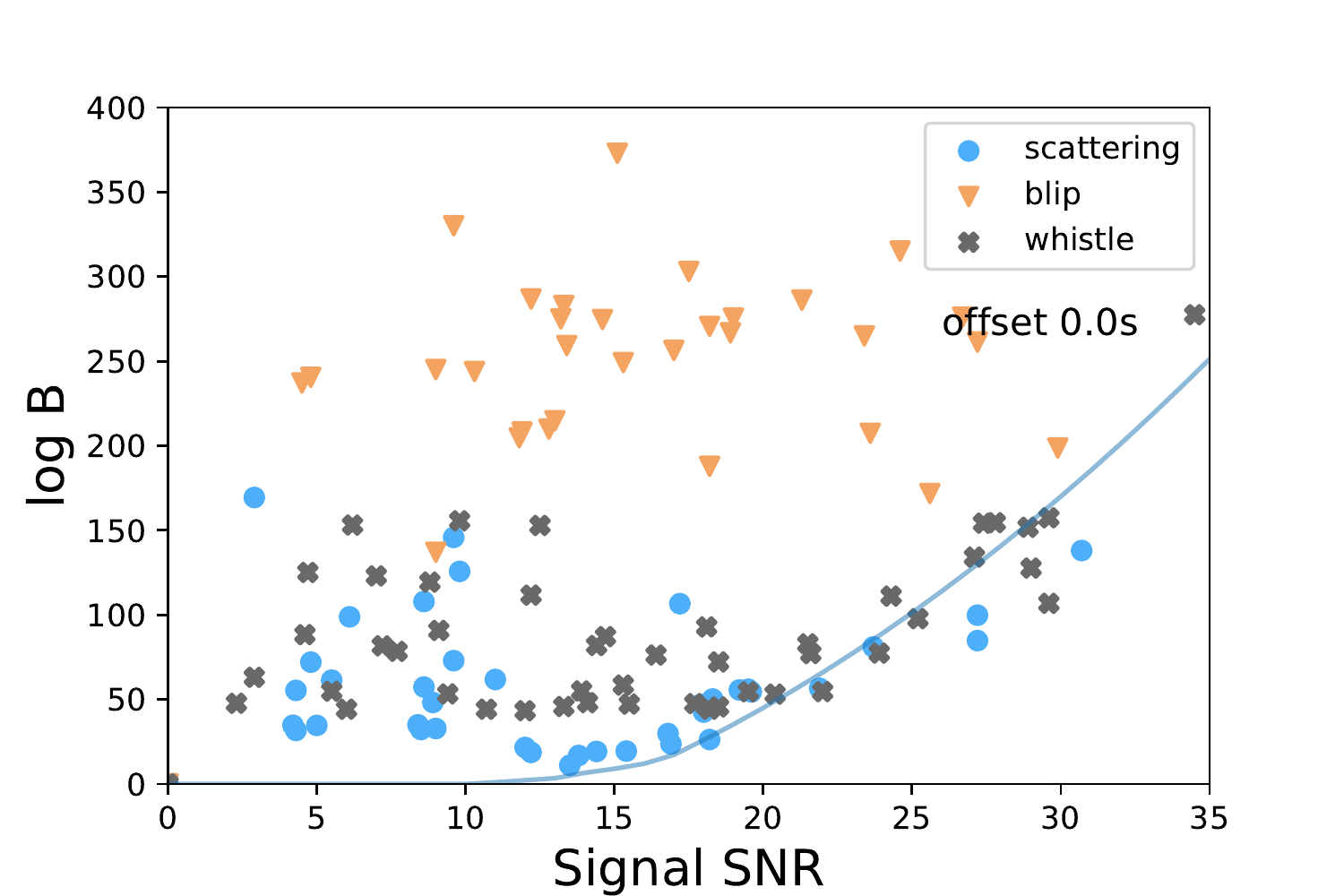}
\includegraphics[width=0.49\textwidth,height=4.5cm]{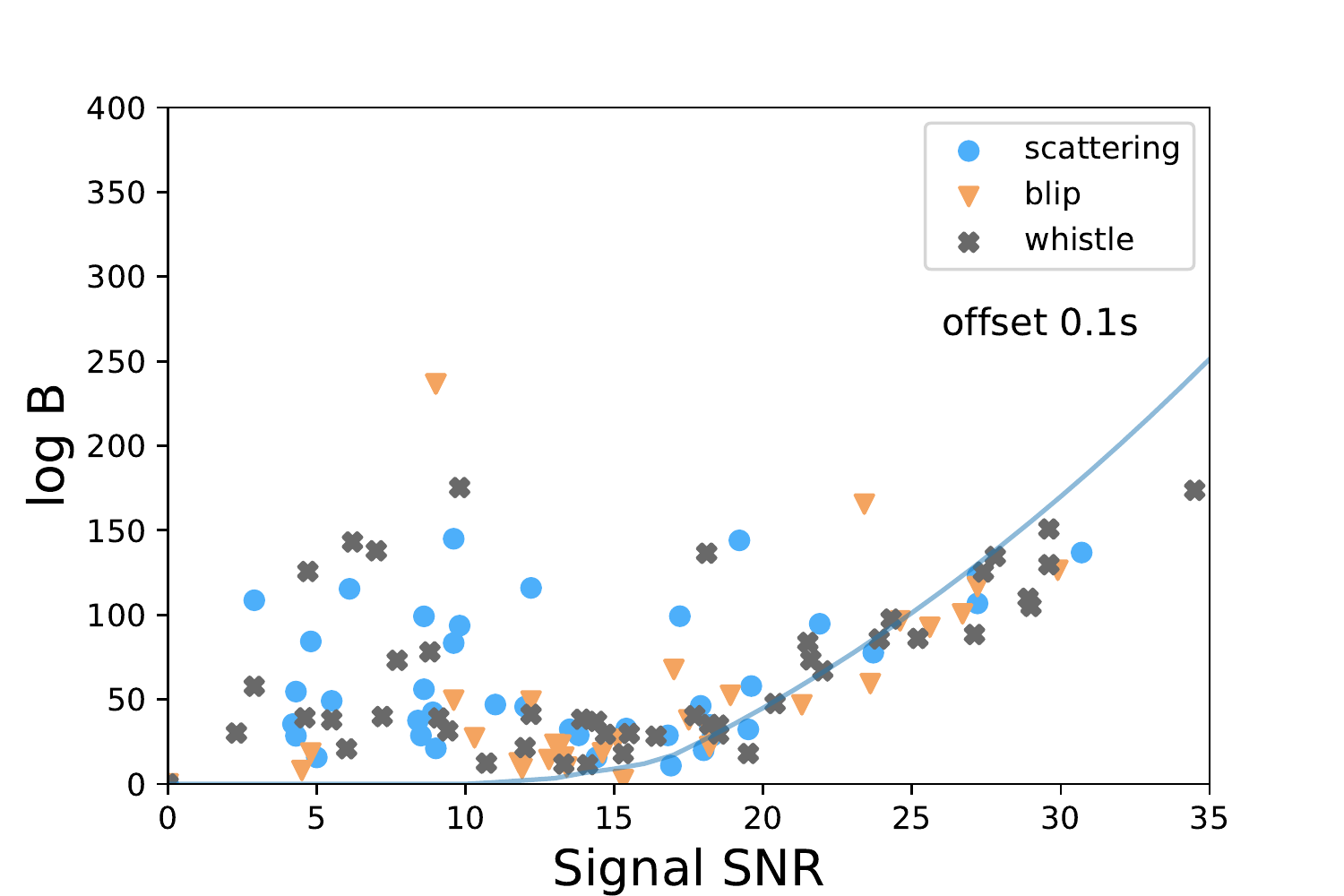}
\includegraphics[width=0.49\textwidth,height=4.5cm]{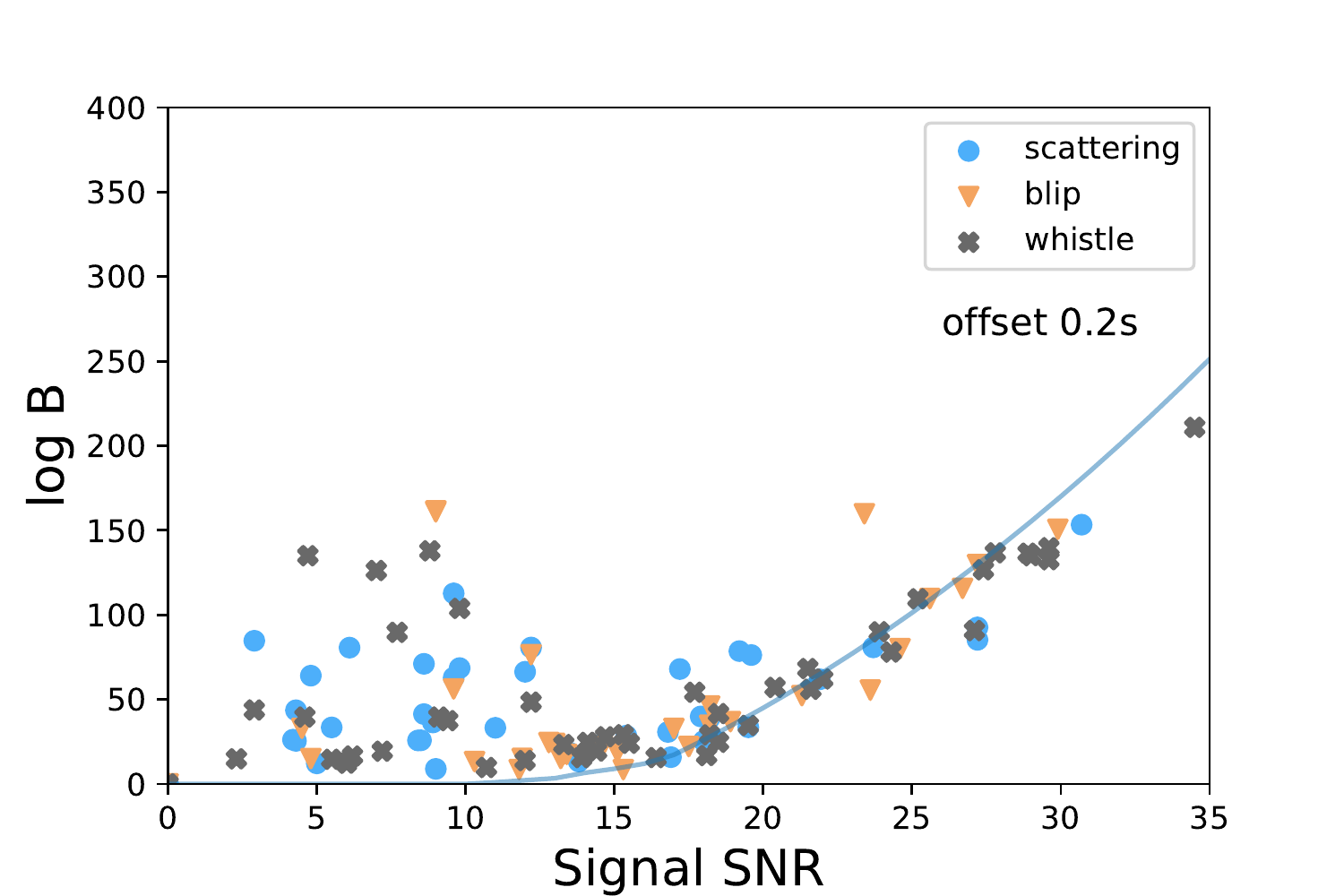}
\includegraphics[width=0.49\textwidth,height=4.5cm]{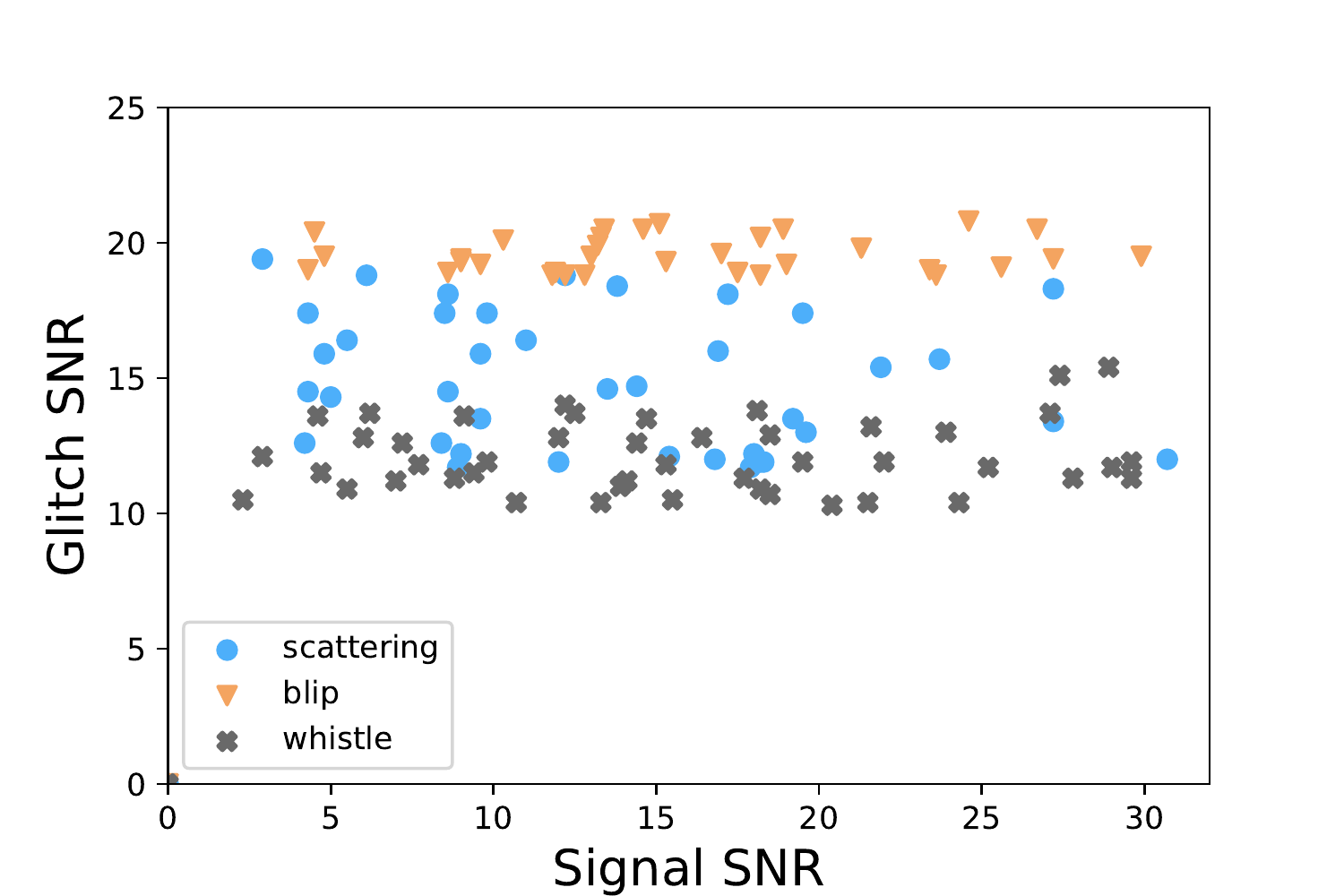}
\caption{ The log Bayes factors for supernova signals with a sine Gaussian model when a glitch is present. The line shows the average expected value when no glitches are present. (Top left) The glitch is directly on top of the signal. (Top right) The glitch is 0.1\,s away from the signal. (Bottom left) The glitch is 0.2\,s away from the signal. (Bottom right) The SNR values for each supernova signal and glitch pair. Blip glitches create the largest increase in the log Bayes factors when the signal is directly on top of the glitch.}
\label{fig:sn_logb}
\end{figure}

In this section, we determine how glitches effect parameter estimation results when there is a mis-match between the signal and the signal model. The injected signals are supernova waveforms, and the signal model is a sine Gaussian. The log Bayes factors, at the three time offsets considered, are shown in Figure \ref{fig:sn_logb}. The mis-match between the signal and model makes the Bayes factors lower than the results for the last two signal types. At all three time offsets considered, the scattered light and whistle glitches increase the log Bayes factors when the SNR of the signal is less than 12. When the signal is directly on top of the glitch, all of the blip glitches create a large increase in the log Bayes factors at all signal SNR values. When the signal is 0.1\,s or 0.2\,s away from the glitch, only one blip glitch increases the log Bayes factor. This is the blip glitch that has the largest SNR. 

\begin{figure}[!t]
\centering
\includegraphics[width=0.48\textwidth,height=4.5cm]{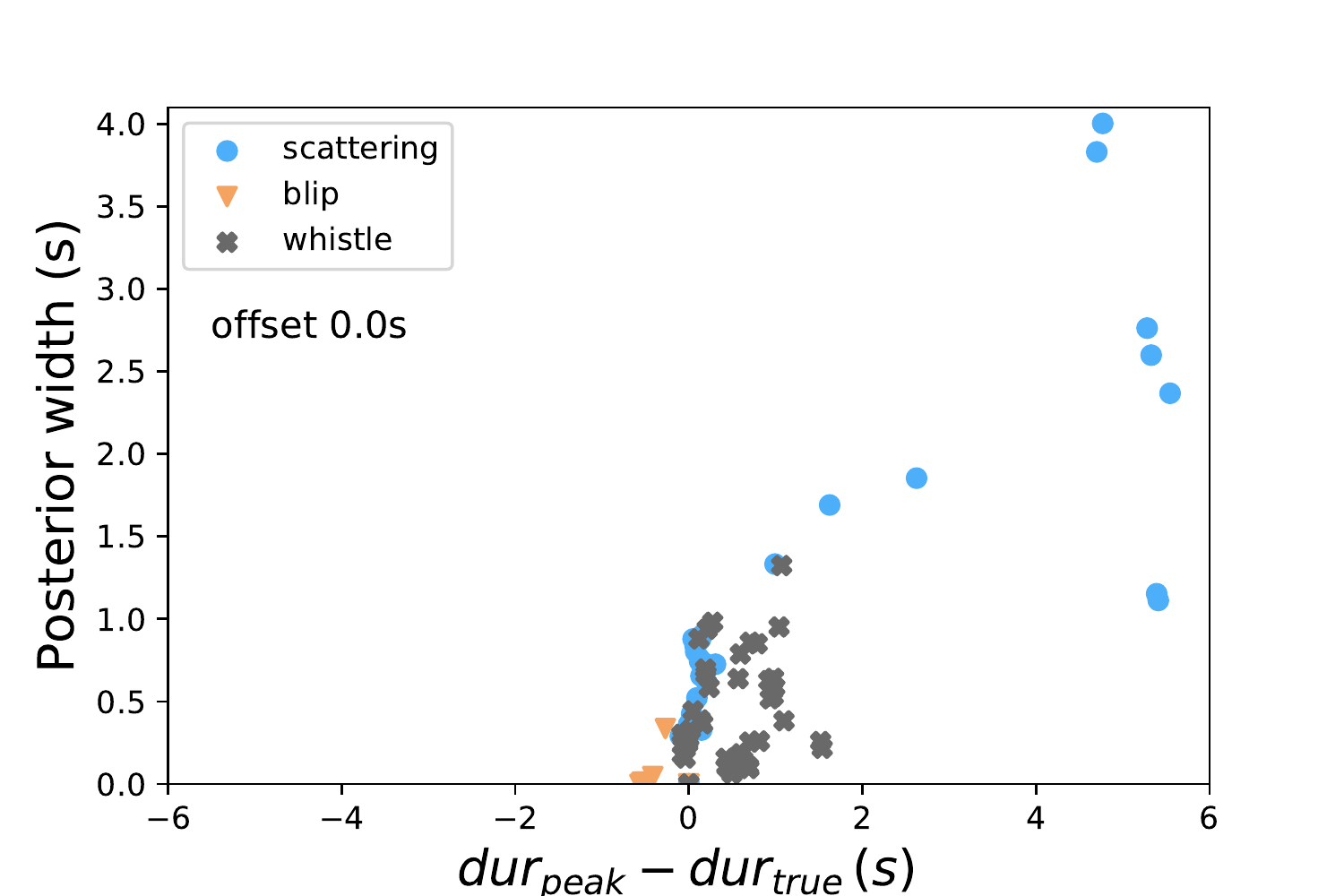}
\includegraphics[width=0.48\textwidth,height=4.5cm]{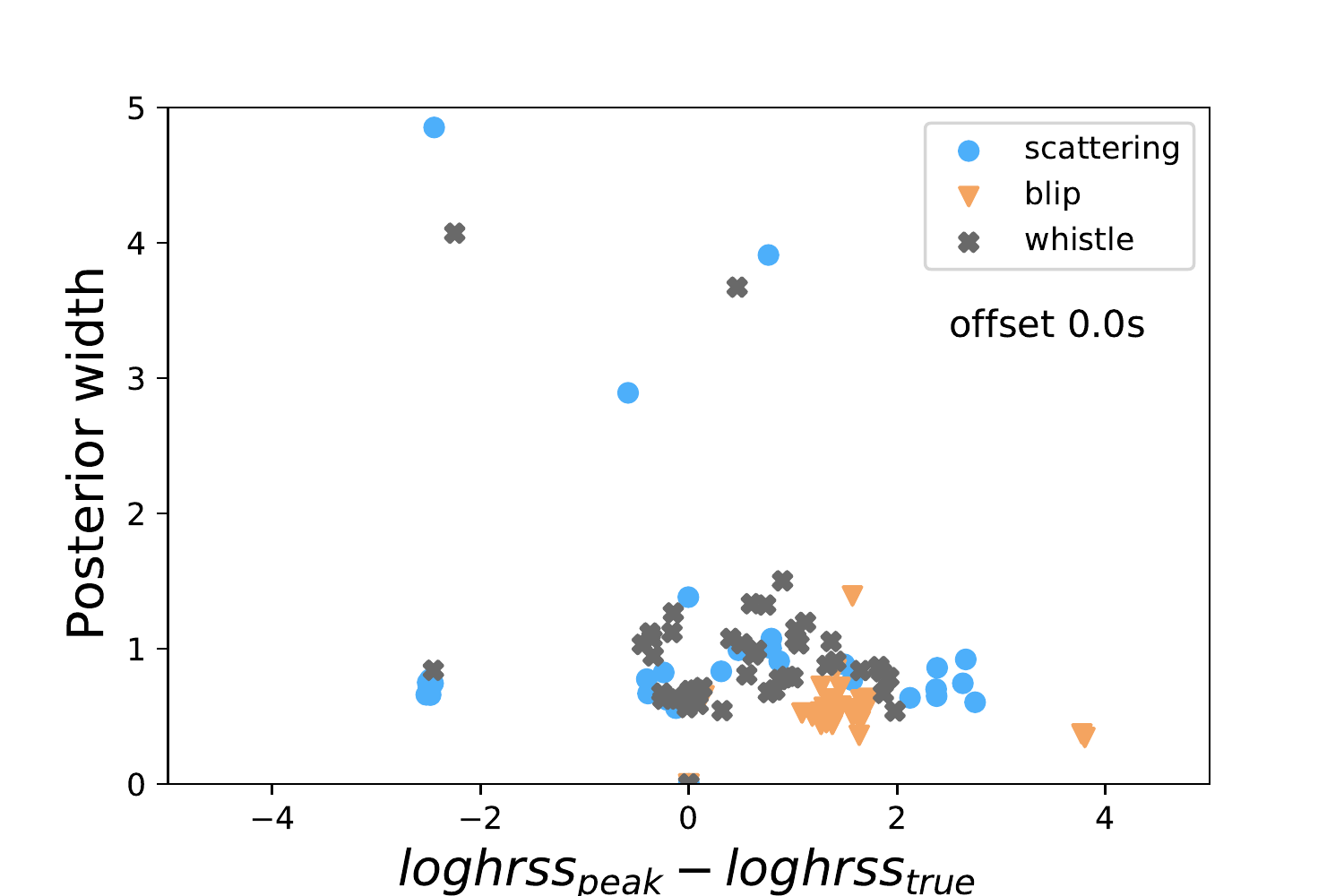}
\includegraphics[width=0.48\textwidth,height=4.5cm]{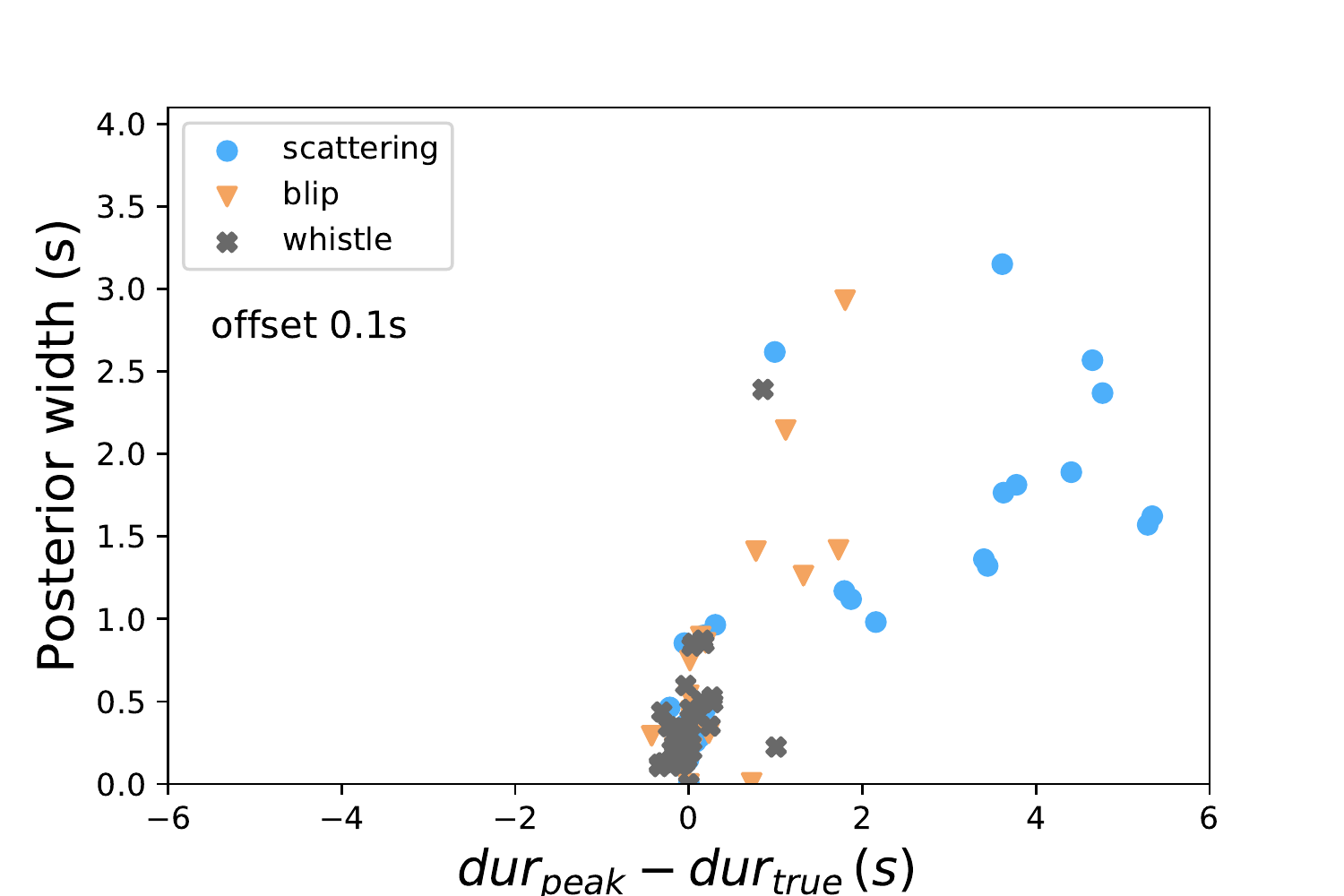}
\includegraphics[width=0.48\textwidth,height=4.5cm]{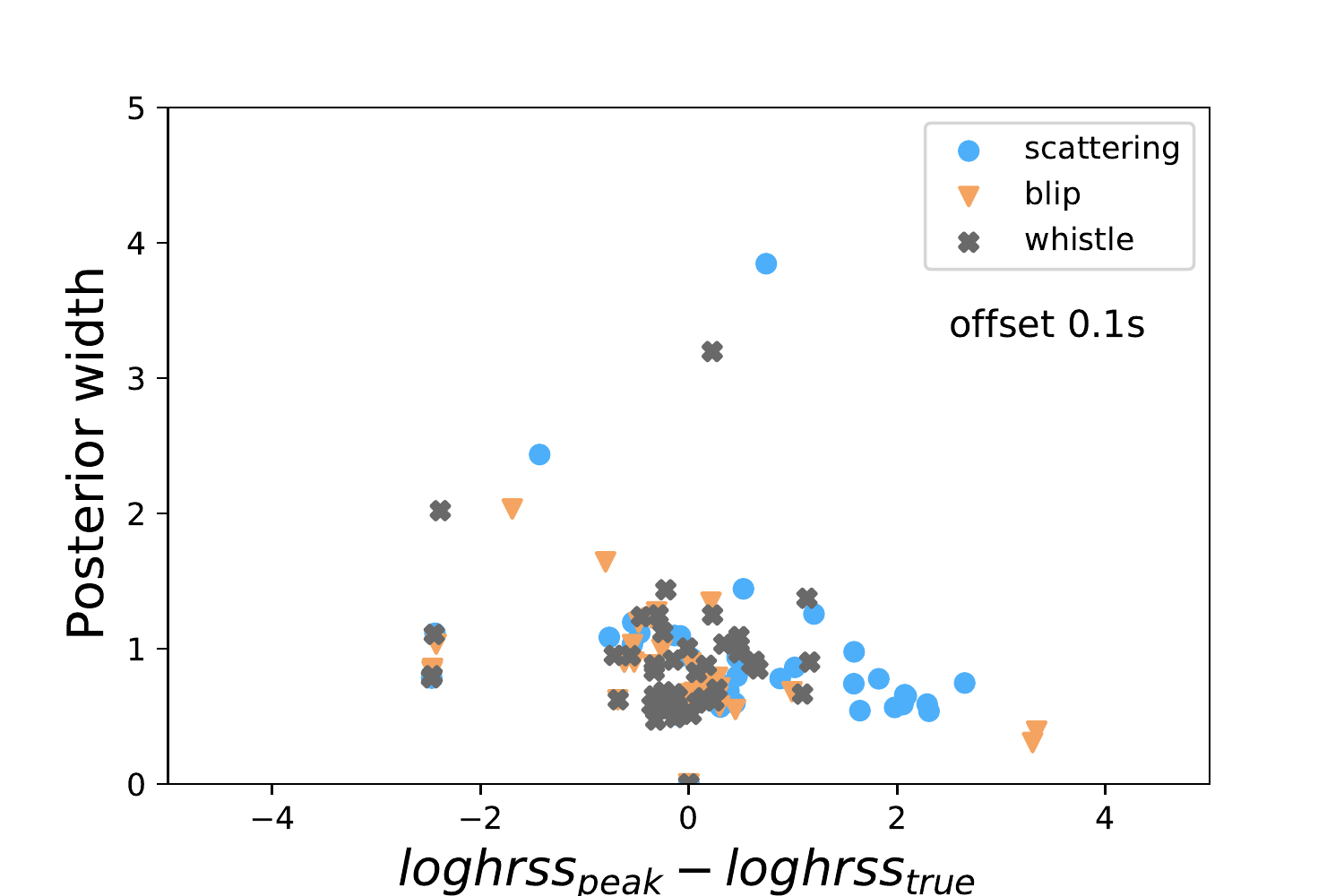}
\includegraphics[width=0.48\textwidth,height=4.5cm]{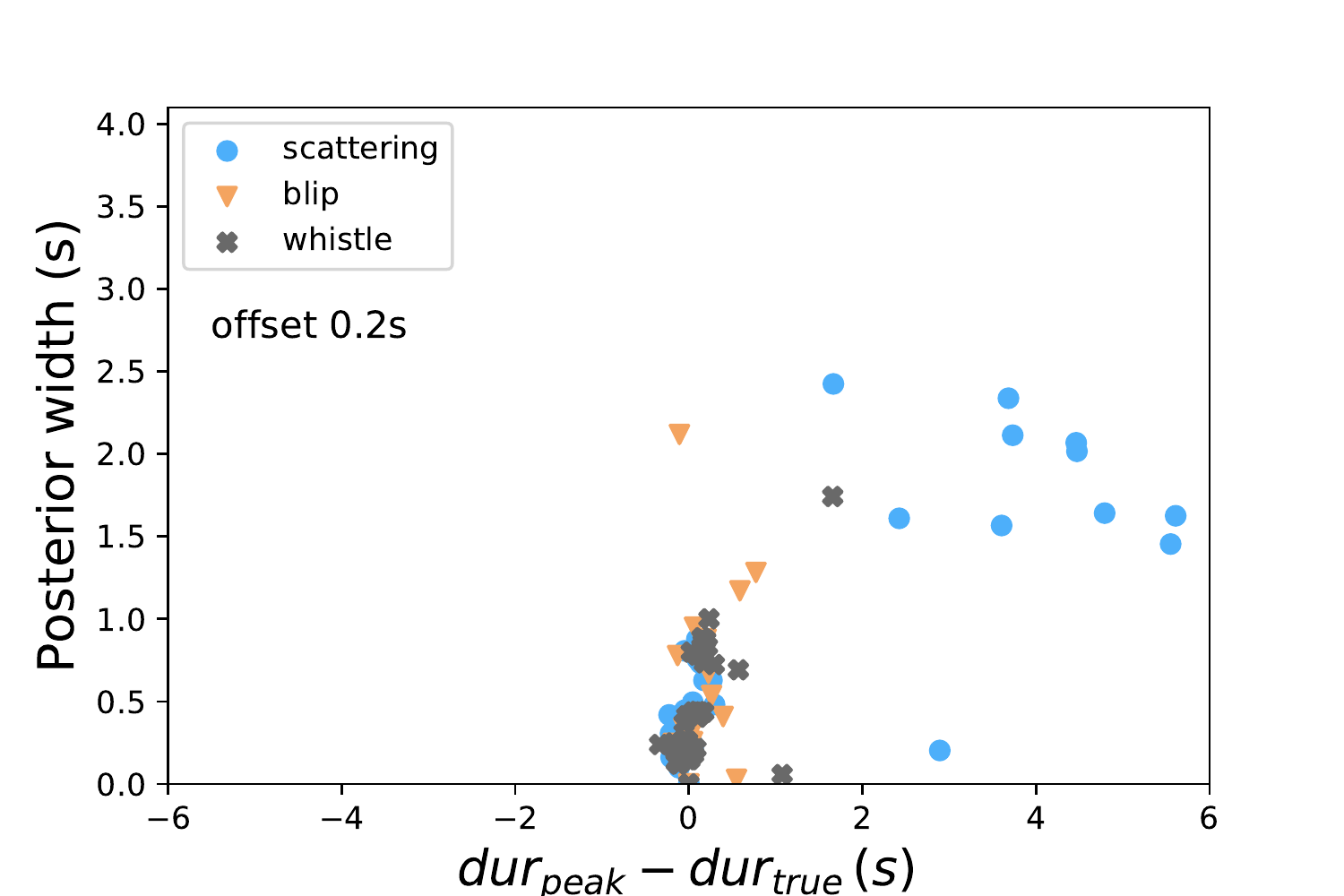}
\includegraphics[width=0.48\textwidth,height=4.5cm]{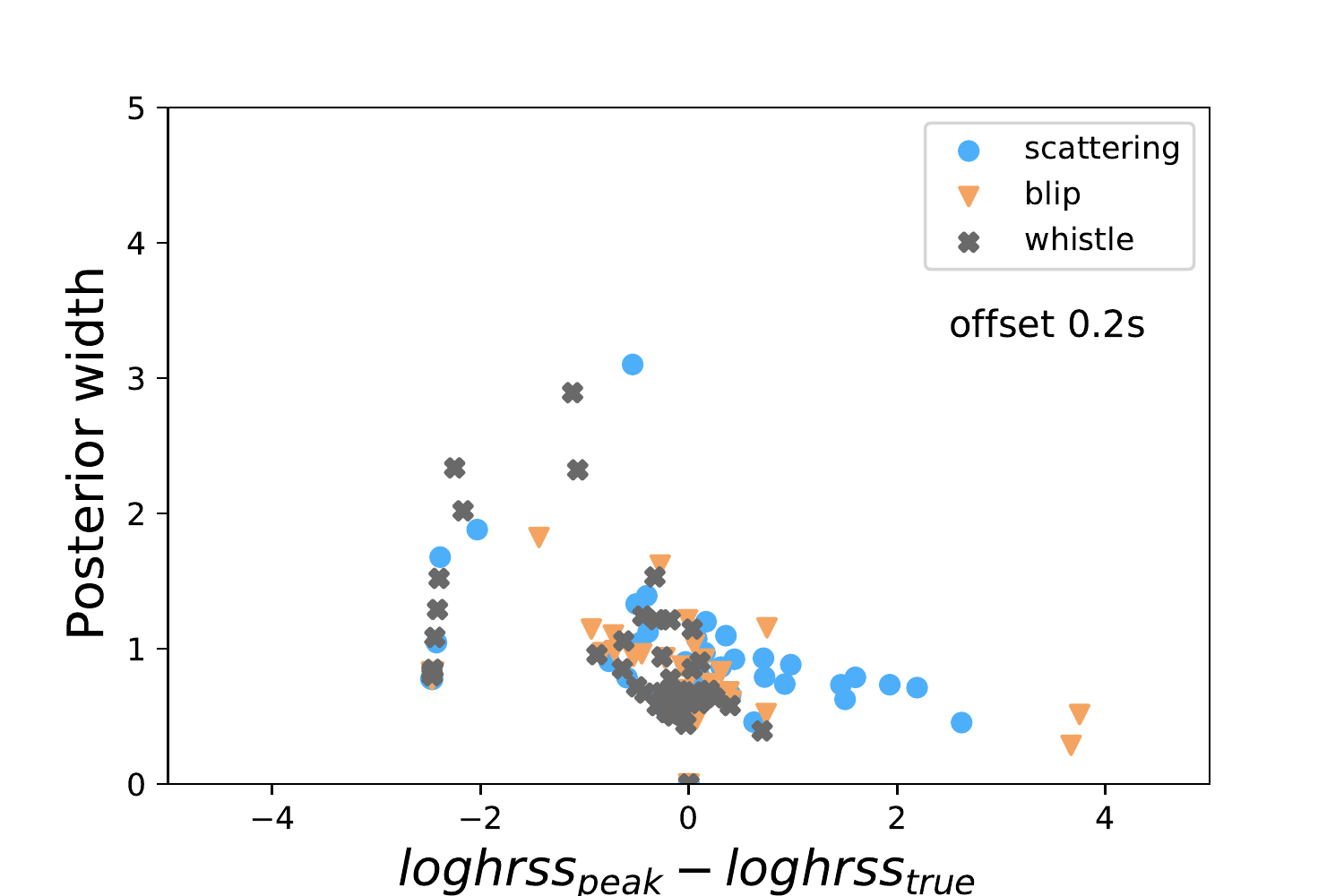}
\caption{For supernova signals and a sine Gaussian signal model, we show the difference between the posterior peak and true duration and hrss parameters, and the width of the 90\% confidence intervals of the posterior distributions. (Top) The glitch is directly on top of the signal. (Middle) The glitch is 0.1\,s away from the signal. (Bottom) The glitch is 0.2\,s away from the signal. Scattered light increases the measured duration of the signals.}
\label{fig:sn_conf}
\end{figure}

The parameter estimation results for the supernova signals are summarized in Figure \ref{fig:sn_conf}. Unlike the results for the other signals, which did not see much effects from the glitches when they are 0.2\,s away from the signal, the supernova results with a mis-match in the template are still effected even when the glitch is 0.2\,s away from the signal. When the signals are directly on top of the glitch, we find that the scattered light glitches make the duration parameter larger than the true value. At 0.2\,s away from the signal, the scattered light glitches still increase the difference between the true duration and the posterior peak, but the posterior width is smaller than when the signal is directly on top of the glitch. We find that all glitch types increase the measured $\log(\mathrm{h}_{rss})$ of the supernova signals. The effect is reduced, but still present when the signal is at larger time offsets from the glitch. 

\subsection{Example supernova posteriors}

\begin{figure}[!t]
\centering
\includegraphics[width=0.49\textwidth,height=4.5cm]{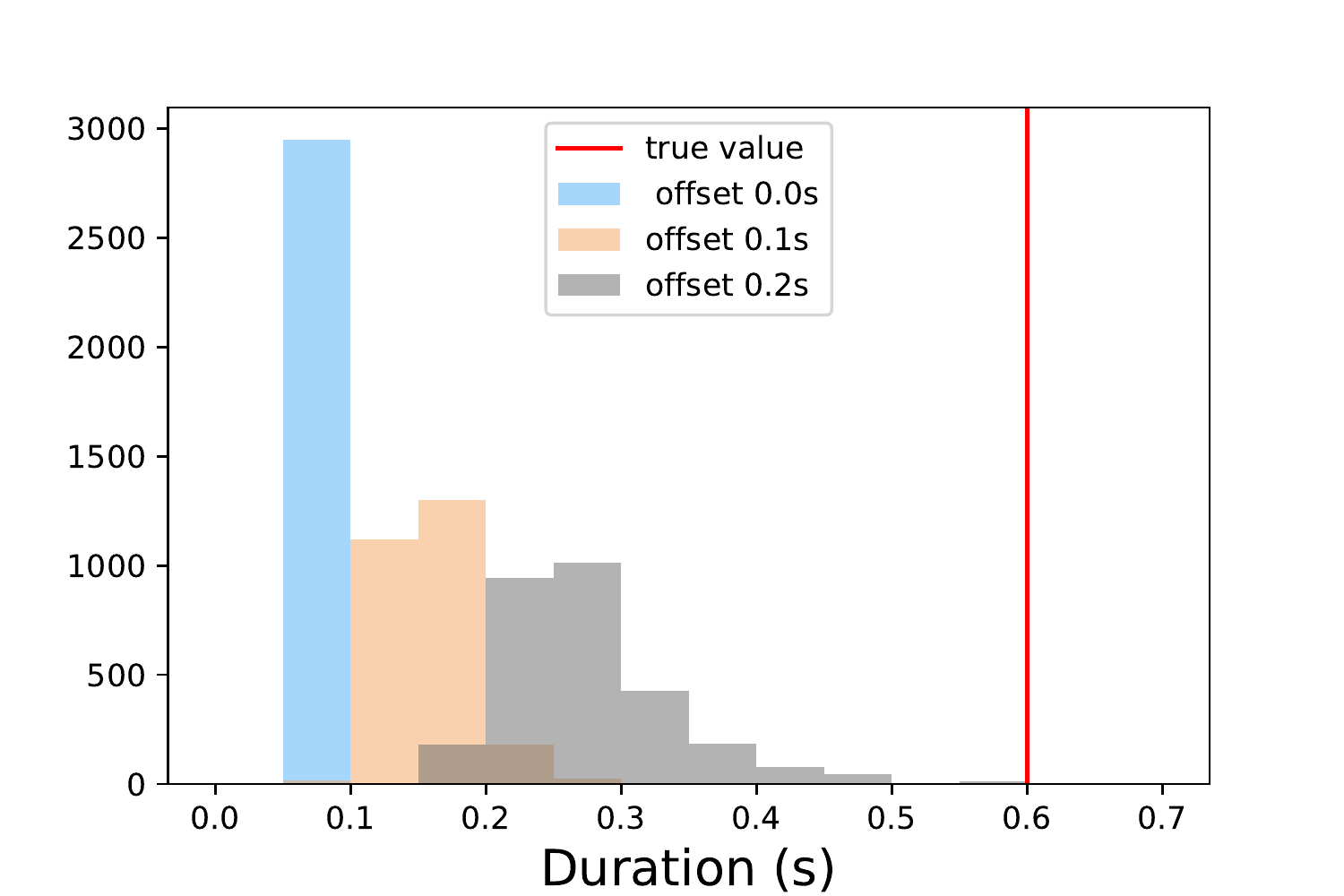}
\includegraphics[width=0.49\textwidth,height=4.5cm]{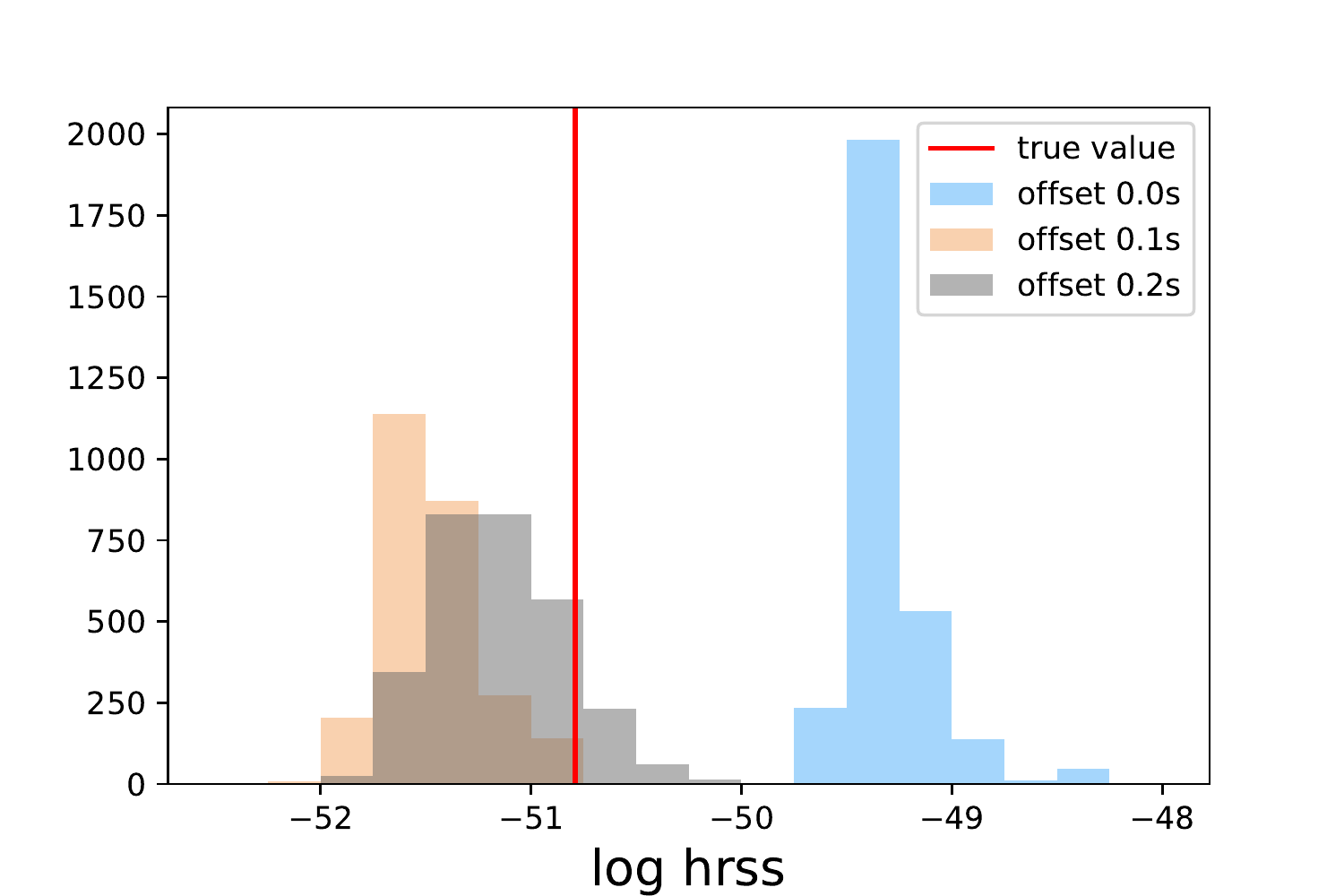}
\caption{An example of the duration and log hrss posteriors for a supernova signal at increasing time offset from a scattered light glitch when there is a mis-match between the signal and model. The posteriors become closer to the true value as the time offset between the signal and glitch increases. }
\label{fig:sn_posts}
\end{figure}

In this subsection, we show in more details the results of one of the supernova signals that was badly affected by a glitch. The example duration and log hrss posteriors are shown in Figure \ref{fig:sn_posts}. The signal is injected on top of a scattered light glitch. The signal network SNR is 12.9, and the glitch SNR is 19.4. In this case, the peak of the posterior distribution for the duration parameter is at a smaller value than expected when the signal is directly on top of the glitch. As the time offset between the signal and glitch becomes larger, the posterior distribution becomes closer to the true value. When the signal is directly on top of the glitch, the measured log hrss posterior distribution is peaked at a larger value than the true value. As the offset between the glitch and the signal becomes larger, the peak of the log hrss and duration posteriors becomes closer to the true value.


\section{Discussion}
\label{sec:discussion}

As gravitational wave detectors become more sensitive, the probability that a detection will overlap with a glitch in one or more of the detectors will increase. Previously when this has occurred it was possible to remove the glitch from the data before any analysis of the signal. This may be more difficult in the future if the signal is short duration, or we don't know exactly what the signal should look like. Therefore, it is important to understand how glitches can effect the results of future detections. 

In this study, we examine the effects of three different types of glitches on the log Bayes factors and estimated parameters of BBH, sine Gaussian and supernova signals at multiple different time offsets from the glitches. Further to this, we examine if the glitches create larger errors when there is a mis-match between the signal and the model. We examine cases where the signal is too short duration for the glitch to be removed, and we include glitches that do not occur in auxiliary channels of data, therefore making it difficult to determine the glitch is present. When the signal and model match, we find that glitches create the largest errors when the signal SNR is smaller than the glitch SNR, and the time offset between signal and glitch is less than 0.1\,s. We find the effect of glitches on the measured signal parameters is worse when there is a mis-match between the signal and model. 

We have shown that for accurate measurements of gravitational wave signals it will be essential in the future to develop further methods to either remove or reduce the effect of the glitches. The gating used for GW170817 is not possible for the signals considered here that are directly on top of the glitches, as their length is either similar to, or smaller than, the glitches considered. Gating glitches near to signals can have negative effects on measurements of the power spectral density (PSD) that is required for the analysis of the signals. Reconstructing the glitch and signal simultaneously may reduce the errors in the parameter measurements of the signals.

Currently it is possible to produce Bayes factors that tell you if the data being analysed contains a signal or a glitch by comparing the results when the data from multiple detectors is analysed coherently and incoherently \cite{PhysRevD.81.062003}. However, there is currently no method to determine a Bayes factor for there being both a signal and glitch at the same time. This is particularly important for signals with unknown waveforms that occur at the same time as glitches that only appear in auxiliary channels of data that are sensitive to gravitational waves, and will be an important area for future development.

\ack
We thank Eric Thrane and Colm Talbot for helpful comments on this work. JP is supported 
by the Australian Research Council Centre of Excellence for Gravitational Wave Discovery 
(OzGrav), through project number CE170100004.


\section*{References}
\bibliographystyle{iopart-num}
\bibliography{bibfile}

\providecommand{\newblock}{}
\begin{thebibliography}{10}
\expandafter\ifx\csname url\endcsname\relax
  \def\url#1{{\tt #1}}\fi
\expandafter\ifx\csname urlprefix\endcsname\relax\def\urlprefix{URL }\fi
\providecommand{\eprint}[2][]{\url{#2}}

\bibitem{aLIGO}
{The LIGO Scientific Collaboration}, {Aasi} J, {Abbott} B~P, {Abbott} R and
  et~al 2015 {\em \cqg\/} {\bf 32} 074001 (\textit{Preprint}
  \eprint{1411.4547})

\bibitem{AdVirgo}
{Acernese, F et al} 2015 {\em \cqg\/} {\bf 32} 024001 (\textit{Preprint}
  \eprint{1408.3978})

\bibitem{PhysRevLett.116.061102}
Abbott B~P {\em et~al.\/} (LIGO Scientific Collaboration and Virgo
  Collaboration) 2016 {\em Phys. Rev. Lett.\/} {\bf 116}(6) 061102
  \urlprefix\url{https://link.aps.org/doi/10.1103/PhysRevLett.116.061102}

\bibitem{PhysRevLett.116.241103}
Abbott B~P, Abbott R {\em et~al.\/} (LIGO Scientific Collaboration and Virgo
  Collaboration) 2016 {\em Phys. Rev. Lett.\/} {\bf 116}(24) 241103
  \urlprefix\url{https://link.aps.org/doi/10.1103/PhysRevLett.116.241103}

\bibitem{PhysRevLett.118.221101}
Abbott B~P, Abbott R {\em et~al.\/} (LIGO Scientific and Virgo Collaboration)
  2017 {\em Phys. Rev. Lett.\/} {\bf 118}(22) 221101
  \urlprefix\url{https://link.aps.org/doi/10.1103/PhysRevLett.118.221101}

\bibitem{2017arXiv171105578T}
{Abbott} B~P, {Abbott} R, {Abbott} T~D, {Acernese} F, {Ackley} K, {Adams} C,
  {Adams} T, {Addesso} P, {Adhikari} R~X, {Adya} V~B and et~al 2017 {\em
  \apjl\/} {\bf 851} L35

\bibitem{PhysRevLett.119.141101}
Abbott B~P, Abbott R {\em et~al.\/} (LIGO Scientific Collaboration and Virgo
  Collaboration) 2017 {\em Phys. Rev. Lett.\/} {\bf 119}(14) 141101
  \urlprefix\url{https://link.aps.org/doi/10.1103/PhysRevLett.119.141101}

\bibitem{PhysRevLett.119.161101}
Abbott B~P {\em et~al.\/} 2017 {\em Phys. Rev. Lett.\/} {\bf 119}(16) 161101
  \urlprefix\url{https://link.aps.org/doi/10.1103/PhysRevLett.119.161101}

\bibitem{2016ApJ...833L...1A}
{Abbott} B~P, {Abbott} R, {Abbott} T~D, {Abernathy} M~R, {Acernese} F, {Ackley}
  K, {Adams} C, {Adams} T, {Addesso} P, {Adhikari} R~X and et~al 2016 {\em
  \apjl\/} {\bf 833} L1 (\textit{Preprint} \eprint{1602.03842})

\bibitem{2016arXiv160501785A}
{Abbott} B~P {\em et~al.\/} (LIGO Scientific Collaboration and Virgo
  Collaboration) 2016 {\em Phys. Rev. D\/} {\bf 94}(10) 102001
  \urlprefix\url{http://link.aps.org/doi/10.1103/PhysRevD.94.102001}

\bibitem{2017arXiv170201759M}
{McNeill} L~O, {Thrane} E and {Lasky} P~D 2017 {\em ArXiv e-prints\/}
  (\textit{Preprint} \eprint{1702.01759})

\bibitem{2005PhRvD..71f3510D}
{Damour} T and {Vilenkin} A 2005 {\em \prd\/} {\bf 71} 063510
  (\textit{Preprint} \eprint{hep-th/0410222})

\bibitem{2041-8205-848-2-L12}
Abbott B~P, Abbott R {\em et~al.\/} 2017 {\em The Astrophysical Journal
  Letters\/} {\bf 848} L12
  \urlprefix\url{http://stacks.iop.org/2041-8205/848/i=2/a=L12}

\bibitem{0264-9381-27-11-114007}
Mandel I and O'Shaughnessy R 2010 {\em Classical and Quantum Gravity\/} {\bf
  27} 114007 \urlprefix\url{http://stacks.iop.org/0264-9381/27/i=11/a=114007}

\bibitem{2017Natur.548..426F}
{Farr} W~M, {Stevenson} S, {Miller} M~C, {Mandel} I, {Farr} B and {Vecchio} A
  2017 {\em Nature\/} {\bf 548} 426--429 (\textit{Preprint}
  \eprint{1706.01385})

\bibitem{2017MNRAS.471.2801S}
{Stevenson} S, {Berry} C~P~L and {Mandel} I 2017 {\em \mnras\/} {\bf 471}
  2801--2811 (\textit{Preprint} \eprint{1703.06873})

\bibitem{2015ApJ...810...58S}
{Stevenson} S, {Ohme} F and {Fairhurst} S 2015 {\em \apj\/} {\bf 810} 58
  (\textit{Preprint} \eprint{1504.07802})

\bibitem{0264-9381-34-3-03LT01}
Vitale S, Lynch R, Sturani R and Graff P 2017 {\em Classical and Quantum
  Gravity\/} {\bf 34} 03LT01
  \urlprefix\url{http://stacks.iop.org/0264-9381/34/i=3/a=03LT01}

\bibitem{2017PhRvD..95l4046G}
{Gerosa} D and {Berti} E 2017 {\em \prd\/} {\bf 95} 124046 (\textit{Preprint}
  \eprint{1703.06223})

\bibitem{2018arXiv180102699T}
{Talbot} C and {Thrane} E 2018 {\em \apj\/} {\bf 856} 173 (\textit{Preprint}
  \eprint{1801.02699})

\bibitem{2015MNRAS.450..414F}
{Fuller} J, {Klion} H, {Abdikamalov} E and {Ott} C~D 2015 {\em \mnras\/} {\bf
  450} 414--427 (\textit{Preprint} \eprint{1501.06951})

\bibitem{2015ApJ...811...86Y}
{Yokozawa} T, {Asano} M, {Kayano} T, {Suwa} Y, {Kanda} N, {Koshio} Y and
  {Vagins} M~R 2015 {\em \apj\/} {\bf 811} 86 (\textit{Preprint}
  \eprint{1410.2050})

\bibitem{2014PhRvD..90d4001A}
{Abdikamalov} E, {Gossan} S, {DeMaio} A~M and {Ott} C~D 2014 {\em \prd\/} {\bf
  90} 044001 (\textit{Preprint} \eprint{1311.3678})

\bibitem{2017PhRvD..95f3019R}
{Richers} S, {Ott} C~D, {Abdikamalov} E, {O'Connor} E and {Sullivan} C 2017
  {\em \prd\/} {\bf 95} 063019 (\textit{Preprint} \eprint{1701.02752})

\bibitem{2017arXiv170801920T}
{Torres-Forn{\'e}} A, {Cerd{\'a}-Dur{\'a}n} P, {Passamonti} A and {Font} J~A
  2018 {\em \mnras\/} {\bf 474} 5272--5286 (\textit{Preprint}
  \eprint{1708.01920})

\bibitem{2016PhRvD..94l3012P}
{Powell} J, {Gossan} S~E, {Logue} J and {Heng} I~S 2016 {\em \prd\/} {\bf 94}
  123012 (\textit{Preprint} \eprint{1610.05573})

\bibitem{2017PhRvD..96l3013P}
{Powell} J, {Szczepanczyk} M and {Heng} I~S 2017 {\em \prd\/} {\bf 96} 123013
  (\textit{Preprint} \eprint{1709.00955})

\bibitem{2015PhRvD..92f4011E}
{Edwards} M~C, {Meyer} R and {Christensen} N 2015 {\em \prd\/} {\bf 92} 064011
  (\textit{Preprint} \eprint{1506.00185})

\bibitem{2015PhRvD..91h4034L}
{Littenberg} T~B and {Cornish} N~J 2015 {\em \prd\/} {\bf 91} 084034
  (\textit{Preprint} \eprint{1410.3852})

\bibitem{2015CQGra..32m5012C}
{Cornish} N~J and {Littenberg} T~B 2015 {\em Classical and Quantum Gravity\/}
  {\bf 32} 135012 (\textit{Preprint} \eprint{1410.3835})

\bibitem{2018CQGra..35f5010A}
{Abbott} B~P, {Abbott} R, {Abbott} T~D, {Abernathy} M~R, {Acernese} F, {Ackley}
  K, {Adams} C, {Adams} T, {Addesso} P, {Adhikari} R~X and et~al 2018 {\em
  Classical and Quantum Gravity\/} {\bf 35} 065010 (\textit{Preprint}
  \eprint{1710.02185})

\bibitem{2016CQGra..33u5004U}
{Usman} S~A, {Nitz} A~H, {Harry} I~W, {Biwer} C~M, {Brown} D~A, {Cabero} M,
  {Capano} C~D, {Dal Canton} T, {Dent} T, {Fairhurst} S, {Kehl} M~S, {Keppel}
  D, {Krishnan} B, {Lenon} A, {Lundgren} A, {Nielsen} A~B, {Pekowsky} L~P,
  {Pfeiffer} H~P, {Saulson} P~R, {West} M and {Willis} J~L 2016 {\em Classical
  and Quantum Gravity\/} {\bf 33} 215004 (\textit{Preprint}
  \eprint{1508.02357})

\bibitem{essick:15}
{Essick} R, {Vitale} S, {Katsavounidis} E, {Vedovato} G and {Klimenko} S 2015
  {\em \apj\/} {\bf 800} 81 (\textit{Preprint} \eprint{1409.2435})

\bibitem{2016PhRvD..93d2004K}
{Klimenko} S {\em et~al.\/} 2016 {\em \prd\/} {\bf 93} 042004
  (\textit{Preprint} \eprint{1511.05999})

\bibitem{2015PhRvD..91d2003V}
{Veitch} J, {Raymond} V, {Farr} B, {Farr} W, {Graff} P, {Vitale} S, {Aylott} B,
  {Blackburn} K, {Christensen} N, {Coughlin} M, {Del Pozzo} W, {Feroz} F,
  {Gair} J, {Haster} C~J, {Kalogera} V, {Littenberg} T, {Mandel} I,
  {O'Shaughnessy} R, {Pitkin} M, {Rodriguez} C, {R{\"o}ver} C, {Sidery} T,
  {Smith} R, {Van Der Sluys} M, {Vecchio} A, {Vousden} W and {Wade} L 2015 {\em
  \prd\/} {\bf 91} 042003 (\textit{Preprint} \eprint{1409.7215})

\bibitem{skilling:04}
{Skilling} J 2004 {Nested Sampling} {\em American Institute of Physics
  Conference Series\/} ({\em American Institute of Physics Conference Series\/}
  vol 735) ed {Fischer} R, {Preuss} R and {Toussaint} U~V pp 395--405

\bibitem{2016PhRvD..93d4006H}
{Husa} S, {Khan} S, {Hannam} M, {P{\"u}rrer} M, {Ohme} F, {Forteza} X~J and
  {Boh{\'e}} A 2016 {\em \prd\/} {\bf 93} 044006 (\textit{Preprint}
  \eprint{1508.07250})

\bibitem{2016PhRvD..93d4007K}
{Khan} S, {Husa} S, {Hannam} M, {Ohme} F, {P{\"u}rrer} M, {Forteza} X~J and
  {Boh{\'e}} A 2016 {\em \prd\/} {\bf 93} 044007 (\textit{Preprint}
  \eprint{1508.07253})

\bibitem{2014PhRvL.113o1101H}
{Hannam} M, {Schmidt} P, {Boh{\'e}} A, {Haegel} L, {Husa} S, {Ohme} F,
  {Pratten} G and {P{\"u}rrer} M 2014 {\em Physical Review Letters\/} {\bf 113}
  151101 (\textit{Preprint} \eprint{1308.3271})

\bibitem{2017arXiv171200688S}
{Smith} R and {Thrane} E 2018 {\em Physical Review X\/} {\bf 8} 021019
  (\textit{Preprint} \eprint{1712.00688})

\bibitem{2016ApJ...829L..15S}
{Singer} L~P, {Chen} H~Y, {Holz} D~E, {Farr} W~M, {Price} L~R, {Raymond} V,
  {Cenko} S~B, {Gehrels} N, {Cannizzo} J, {Kasliwal} M~M, {Nissanke} S,
  {Coughlin} M, {Farr} B, {Urban} A~L, {Vitale} S, {Veitch} J, {Graff} P,
  {Berry} C~P~L, {Mohapatra} S and {Mandel} I 2016 {\em \apjl\/} {\bf 829} L15
  (\textit{Preprint} \eprint{1603.07333})

\bibitem{2017Natur.551...85A}
{Abbott} B~P, {Abbott} R, {Abbott} T~D, {Acernese} F, {Ackley} K, {Adams} C,
  {Adams} T, {Addesso} P, {Adhikari} R~X, {Adya} V~B and et~al 2017 {\em
  Nature\/} {\bf 551} 85--88 (\textit{Preprint} \eprint{1710.05835})

\bibitem{lynch:15}
{Lynch} R, {Vitale} S, {Essick} R, {Katsavounidis} E and {Robinet} F 2015 {\em
  arXiv:1511.05955\/} (\textit{Preprint} \eprint{1511.05955})

\bibitem{gravityspy}
{Zevin} M, {Coughlin} S, {Bahaadini} S, {Besler} E, {Rohani} N, {Allen} S,
  {Cabero} M, {Crowston} K, {Katsaggelos} A~K, {Larson} S, {Lee} T~K, {Lintott}
  C, {Littenberg} T~B, {Lundgren} A, {Osterlund} C, {Smith} J~R, {Trouille} L
  and {Kalogera} V 2017 {\em Classical and Quantum Gravity\/} {\bf 34} 064003
  \urlprefix\url{http://stacks.iop.org/0264-9381/34/i=6/a=064003}

\bibitem{2017CQGra..34c4002P}
{Powell} J, {Torres-Forn{\'e}} A, {Lynch} R, {Trifir{\`o}} D, {Cuoco} E,
  {Cavagli{\`a}} M, {Heng} I~S and {Font} J~A 2017 {\em Classical and Quantum
  Gravity\/} {\bf 34} 034002 (\textit{Preprint} \eprint{1609.06262})

\bibitem{2015CQGra..32u5012P}
{Powell} J, {Trifir{\`o}} D, {Cuoco} E, {Heng} I~S and {Cavagli{\`a}} M 2015
  {\em Classical and Quantum Gravity\/} {\bf 32} 215012 (\textit{Preprint}
  \eprint{1505.01299})

\bibitem{nuttall:15}
{Nuttall} L~K {\em et~al.\/} 2015 {\em Classical and Quantum Gravity\/} {\bf
  32} 245005 (\textit{Preprint} \eprint{1508.07316})

\bibitem{mueller:e12}
{M{\"u}ller} E, {Janka} H~T and {Wongwathanarat} A 2012 {\em \aap\/} {\bf 537}
  A63 (\textit{Preprint} \eprint{1106.6301})

\bibitem{PhysRevD.81.062003}
Veitch J and Vecchio A 2010 {\em Phys. Rev. D\/} {\bf 81}(6) 062003
  \urlprefix\url{https://link.aps.org/doi/10.1103/PhysRevD.81.062003}

\end{thebibliography}

\end{document}